\documentclass[sn-mathphys,Numbered]{sn-jnl}

\usepackage{graphicx}%
\usepackage{multirow}%
\usepackage{amsmath,amssymb,amsfonts}%
\usepackage{amsthm}%
\usepackage{mathrsfs}%
\usepackage[title]{appendix}%
\usepackage{xcolor}%
\usepackage{textcomp}%
\usepackage{manyfoot}%
\usepackage{booktabs}%
\usepackage{algorithm}%
\usepackage{algorithmicx}%
\usepackage{algpseudocode}%
\usepackage{listings}%

\usepackage{xspace}
\newcommand{\ie}{\emph{i.e.,}\xspace}
\newcommand{\eg}{\emph{e.g.,}\xspace}
\newcommand{\etc}{etc.\xspace}

\usepackage{tabularx}
\usepackage{pifont} 

\usepackage{tcolorbox} 
\usepackage{adjustbox} 
\usepackage{multirow}
\usepackage{graphicx}
\usepackage{xcolor,colortbl}
\definecolor{gray10}{gray}{.9}

\usepackage{subcaption} 

\usepackage{cleveref}

\usepackage{rotating}

\usepackage{float}

\usepackage [autostyle, english = american]{csquotes}
\MakeOuterQuote{"}

\begin{document}

\title[Article Title]{The Sustainability Assessment Framework Toolkit: A Decade of Modeling Experience}


\author*[1]{\fnm{Patricia} \sur{Lago}}\email{p.lago@vu.nl}

\author[2]{\fnm{Nelly} \sur{Condori Fernandez}}\email{n.condori.fernandez@usc.es}
\equalcont{These authors contributed equally to this work.}

\author[1]{\fnm{Iffat} \sur{Fatima}}\email{i.fatima@vu.nl}
\equalcont{These authors contributed equally to this work.}

\author[1]{\fnm{Markus} \sur{Funke}}\email{m.t.funke@vu.nl}
\equalcont{These authors contributed equally to this work.}

\author[1]{\fnm{Ivano} \sur{Malavolta}}\email{i.malavolta@vu.nl}
\equalcont{These authors contributed equally to this work.}

\affil[1]{\orgdiv{Software and Sustainability (S2) Research Group}, \orgname{Vrije Universiteit Amsterdam}, \country{The Netherlands}}

\affil[2]{\orgdiv{Centro Singular de Investigación en Tecnoloxías Intelixentes (CiTIUS)}, \orgname{University of Santiago de Compostela}, \country{Spain}}



\abstract{Software intensive systems play a crucial role in most, if not all, aspects of modern society. As such, both their sustainability and their role in supporting sustainable processes must be realized \textit{by design}. To this aim, the architecture of software intensive systems should be designed to support sustainability goals; and measured to understand how effectively they do so. 

In this paper, we present the Sustainability Assessment Framework (SAF) Toolkit -- a set of instruments we developed to support software architects and design decision makers in modeling sustainability as a software quality property. The SAF Toolkit is the result of our experience gained in more than a decade of case studies in collaboration with industrial partners. We illustrate the toolkit with examples that come from some of such studies. We extract our lessons learned, our current research, and future plans to extend the SAF Toolkit for further architecture modeling and measurement.}




\maketitle


\section{Introduction}\label{sec:introduction}

Software-intensive systems play a crucial role in most, if not all, aspects of modern society. As such, both their own sustainability and their role in supporting sustainable processes must be realized \textit{by design}.

The Energy Efficiency Directive for Data Centers~\cite{EED2023}, reports that \textit{in the EU, from 2010 to 2018 data centre energy consumption increased by 42\% and is forecast to further increase by 28.2\% by 2030, representing about 3.2\% of  the  EU  final  electricity demand.  Reducing  the  energy  demand  of  ICT,  including  data  centres,  is  an important step in the target reduction of overall GHG emissions of 55\% by 2030 compared to 1990 levels}. Given that nowadays most software is to various degrees cloud based (\eg~\cite{cloudsoftware2023a, cloudsoftware2023b}), its energy efficiency is imperative~\cite{Moghaddam2015}. If we also consider the global adoption of AI-based solutions like generative AI~\cite{generative_ai}, there is growing evidence that the energy footprint of software will escalate in the coming years~\cite{verdecchia2023systematic}.

Next to the environmental perspective, sustainability entails other dimensions which further increase the role of software in society at large: from a social perspective software usability, for example, can support or hinder accessibility to fundamental services like healthcare and education; from an economic perspective, affordability of software products and related technologies can significantly influence economic growth; and of course from a technical perspective, evolvability or integrability of software to accommodate over time the changes in society and consumer needs, can act as the motor for, or against, innovation.

As sustainability and its dimensions have a broad scope and are by and large interdependent, software professionals and decision makers need to consider software sustainability at a high level of abstraction. This way, the role of software elements and the way they influence one another can be (re-)designed for, monitored, and reasoned about. We believe that the software architecture level provides the right abstraction needed for the above.

The above-mentioned problem can be framed in the following overarching research question (RQ): \textit{How can we support software architects and design decision makers in modeling sustainability as a software quality property?}
In this paper, we present the Sustainability Assessment Framework (SAF) Toolkit -- a set of instruments that aim to address this RQ. 

The \textbf{intended target audience} of this paper is (i) practitioners responsible for both setting priorities to bring software sustainability to the forefront of their organizations and deciding how sustainability concerns are designed in software solutions, and (ii) instructors willing to include software sustainability into Computer Science and Software Engineering programs. Both can access the SAF Toolkit guidelines, tutorials, and links to open-source instruments in the \textbf{supplementary online material}~\cite{saf_toolkit_online}.

This work builds on our previous publication \cite{Condori-FernandezEtAl_ActionResearch_2020}, where we first introduced the SAF instruments and reported the results of an action-research initiative aimed at improving them. In this paper, we offer an exhaustive description of each refined instrument (i.e., the Decision Map and the Sustainability-Quality model) by illustrating the toolkit with examples from various case studies (the full list of related cases is provided in \Cref{tab:cases}).
Furthermore, we share our plans to extend the SAF Toolkit with a Key Performance Indicator (KPI) model and the complete SAF Toolkit workflow. These extensions aim to contribute to a more robust and effective sustainability assessment from a software architecture perspective.
This work represents a significant evolution of our approach, showcasing the practical application and improvements achieved over a decade of modeling experience. 

In summary, the \textbf{main contributions} of this paper are:
\begin{itemize}
    \item a complete, overarching presentation of \textit{the SAF Toolkit} with all its guiding instruments, the related illustrative examples and reflections explaining how and why the instruments evolved thanks to the series of case studies carried out over the years;
    \item an \textit{overview of the case studies} carried out in the last decade, with pointers to related publications;
    \item the description of the \textit{future extensions} to the SAF Toolkit, namely (i) the integration of KPIs for measuring and monitoring how implemented architectures address targeted sustainability goals; (ii) the integration of software architecture descriptions for full traceability of which architecture elements are designed to be responsible for which measured sustainability impacts; (iii) the integration with the Green Lab, an experimental platform for conducting empirical studies on the quality of software with a special emphasis on its energy efficiency; and (iv) the addition of a monitoring dashboard that will provide the toolkit users with dynamic insights into runtime measurements and which associated elements in which SAF instruments should possibly be redesigned for sustainability improvements;
    \item a \textit{definition of sustainability} that emphasizes the indivisible inclusion of both dimensions of Sustainability-Quality and of Time.
\end{itemize}

The remainder of this paper is organized as follows.
Section~\ref{sec:background} presents the background on the notion of sustainability, its relation to software as used in our work, and the notion of KPI, while Section~\ref{sec:rw} presents related works.
Section~\ref{sec:toolkit} describes the SAF Toolkit and its instruments in detail, and explains the current release and its evolution over time, and the next steps for future research.
In Section~\ref{sec:SAFinAction} we link all instruments together and outline our vision on how the complete SAF Toolkit workflow can be used in practice.
Finally, in Section~\ref{sec:discussion} we discuss additional overarching observations and conclude the paper.

\section{Background}\label{sec:background}

In this section, we provide the background knowledge necessary for the remainder of this paper. In Section~\ref{sec:def-sust}, we introduce the definition of sustainability in general and specifically in the context of software. We then provide an overview of sustainability from a qualitative and a quantitative perspective. Qualitatively, as illustrated in the right-hand side of Figure~\ref{fig:toolkitModel} (see yellow classes), sustainability involves two indivisible dimensions, 
namely the dimension expressing the various sustainability aspects of a software-intensive system in terms of quality properties (see Section~\ref{sec:Q-dim}), and the dimension expressing the related impact over time (see Section~\ref{sec:T-dim}). Quantitatively, in Section~\ref{sec:KPIs} we use the concept of KPI to express the corresponding software sustainability measures.

\subsection{Definition of Software Sustainability}\label{sec:def-sust}

In almost every sustainability research, regardless of the discipline, the definition of sustainability is repeatedly redefined without finding a common ground, resulting in a wide range of interpretations~\cite{CaleroEtAl_5WsGreen_2020, PetersEtAl_SustainabilityComputing_2024,CaleroPiattini_PuzzlingOut_2017}.
In software sustainability, \citet{Penzenstadler2013-3criteria} proposes definitions to consider three variables, namely system, function and scope, the latter possibly mapped to a time scope; and tries to link these variables to possible interpretations of the known notions of software for sustainability, sustainable software, and sustainable software engineering process. \citet{Wolfram2017-def} review the literature on existing definitions and, among other things, observe that the time scope is largely neglected.

In turn, \citet{VentersEtAl_SustainableSoftware_2023} 
discuss the ``maturity of software sustainability from a software architecture point of view''. The authors confirm that there is no universally accepted definition of software sustainability for either era; software sustainability is either "a measure of endurance" or "aligned [...] to one or more software quality attributes". However, in addition to technical and environmental aspects, the literature now recognizes other dimensions~\cite{LagoEtAl_FramingSustainability_2015} such as the economic, social, and individual ones~\cite{VentersEtAl_SustainableSoftware_2023} (see also \Cref{sec:Q-dim}).

Due to this variety of definitions and the importance of software in an increasingly digitalized society, we see software sustainability as the software "ensuring that the intended benefits are preserved over time, notwithstanding who uses them or where they are deployed"~\cite{Lago_ConnectedWorld_2023}. For a more comprehensive understanding of sustainability, we encourage the interested reader to explore various in-depth examinations and definitions available on the subject,~\eg~\cite{PetersEtAl_SustainabilityComputing_2024,BeckerEtAl_SustainabilityDesign_2015, CaleroEtAl_5WsGreen_2020, VentersEtAl_BlindMen_2014, CaleroEtAl_IntroductionSoftware_2021, AndrikopoulosEtAl_SustainabilitySoftware_2022,Heldal2024-industry}.

\subsection{The Dimension of ``Sustainability Quality'' (or Sustainability-Quality Dimension)} \label{sec:Q-dim}


As stated above, achieving sustainability involves various dimensions and finding a balance between them while also considering their trade-offs~\cite{LagoEtAl_FramingSustainability_2015}. 
At a higher abstraction level, the PESTLE model examines sustainability from Political, Economic, Social, Technological, Legal, and Environmental perspectives to identify challenges and opportunities for businesses~\cite{Nandonde_PESTLEAnalysis_2019}. The ESG framework focuses on a smaller subset of factors, namely Environmental, Social, and corporate Governance, which should be aligned to ensure sustainable investment in the financial market and international organisations~\cite{EuropeanBankingAuthority_ReportIncorporating_2022}.

Also in computer sciences, there are different perspectives on sustainability. 
The SAF with its focus on software architecture in particular reflects on the four sustainability dimensions according to \citet{LagoEtAl_FramingSustainability_2015}: (i) \textit{Technical} dimension encompasses implementation aspects, as well as concerns related to the evolution, maintenance, and long-term utilisation of software-intensive systems; (ii) \textit{Economic} dimension revolves around business considerations, such as capital investment and profitability; (iii) \textit{Social} dimension emphasises the integration of the systems within communities considering their impact on individuals and society; (iv) \textit{Environmental} dimension extends beyond greenhouse gas emissions and encompasses the broader impacts of software activities on the natural ecosystem. In requirements engineering, for instance, a fifth \textit{Individual} dimension has emerged as important for reflecting on an individual's well-being~\cite{DubocEtAl_RequirementsEngineering_2020}.


\subsection{The Dimension of "Time" (or Sustainability-Time Dimension)}\label{sec:T-dim}

In addition to the different sustainability dimensions, sustainability has another facet \ie \textit{time}. \citet{Hilty_2015} present a framework for ICT impacts on sustainability with a three-level model for the over-time effects of ICT: (i) Level 1 refers to the \textit{direct effects} of the production, use or disposal of ICT; (ii) Level 2 refers to the \textit{enabling effects} of ICT as induction, obsolescence, substitution and optimization effects; (iii) Level 3 refers to the\textit{ systemic effects} which are the long-term socio-economic effects that induce a behavioral or economic structural change. This can be viewed as both negative rebound effects and positive cascading effects. \citet{Hilty_2015} also identify an inherent limitation of their framework as it mixes up the levels of abstraction and the categories of effects. It is not clear how micro-level action across one dimension can lead to a macro-level impact without impacting other dimensions of sustainability. Hence, an isolated action cannot be considered to have a strictly positive or negative effect unless its macro-level impact is studied. As sustainability goals take a macro-level perspective, this phenomenon requires a macro-level analysis.

The intended benefits of sustainability decisions must uphold with ever-evolving nature of the digital landscape. In view of this, \citet{Lago_ConnectedWorld_2023} defines sustainability as ``the preservation of the beneficial use of digital solutions, in a context that continuously changes". Accordingly, software must be viewed for its over-time impact on sustainability.

\subsection{Key Performance Indicators} \label{sec:KPIs}

To supplement micro- and macro-level support for sustainability impact assessment across the different levels of abstraction, KPIs can provide a solution by aggregating micro-level \textit{data} to macro-level \textit{indicators} to monitor sustainability concerns. 
\citet{Parmenter_2015} defines a KPI as ``an indicator that focuses on the aspects of organizational performance that are the most critical for the current and future success of the organization" \cite{Parmenter_2015}. To formalize KPIs for software, \citet{Fatima_2023} revise Parmenter's definition to ``an indicator that focuses on the aspects of \textit{software} performance that are the most critical for the current and future success of the organization''. By focusing on software performance aspects and linking it with organizational success, a KPI enables integrating software quality into organizational strategy as an intrinsic mechanism. Thus, KPIs act as a linchpin between the business and the technical ends of digital organizations for monitoring success. It is important to note that the word ‘performance’ in KPI does not refer to the Quality Attribute (QA) performance, but rather it means how well the software satisfies the critical success factors through its operation.

In the context of software, \citet{Staron_2016} define a KPI as ``a critical element in the transformation of raw data (numbers) into decisions (indicators)''. For example, in a later work~\cite{Staron_2012} they present a release readiness KPI to check when a product will be ready for release based on the number of defects present, defects removed, and predicted defect discovery. A decision criterion is established based on thresholds or patterns. If the values deviate from a threshold or pattern, the KPI monitoring enables the quality manager to make decisions to trigger actions for improvement. The same concept can be applied to the field of software architecture by modeling software quality through KPIs, their targets representing the achievement of a design concern, and by using these KPIs to drive architectural design decisions through actions.

\section{Related Work}\label{sec:rw} 

In this section, we discuss the related works that focus on modeling frameworks or assessment initiatives aimed at achieving software sustainability. 

\citet{Naumann_2011} present a reference model for Green and Sustainable Software -- the GREENSOFT model. This model is designed to capture sustainability effects (first-, second-, and third-order of effects) of software from their inception until their end-of-life. The model is supplemented by metrics for sustainability criteria (common, direct and indirect). The model also provides procedure models for the development, purchase, operation and use of software in a sustainably-sound manner. It rounds off by providing recommendations and tools for software development, purchase, administration and usage. The GREENSOFT model focuses on both Green of IT and Green by IT. 

\citet{Cabot_2009} use the i* framework for the integration of sustainability design concerns in (i) design alternative selection for software, and (ii) decisions for business activities. The use of this framework is demonstrated through a case study by identifying sustainability-specific soft goals, tasks, agents and roles. Dependencies and relationships are identified between these elements. Goals are linked to each other with the relation of positive (`help') or negative (`hurt') impacts. The study acknowledges the challenges of the proposed approach as (i) difficulty in determining the appropriate level of detail for effectively representing sustainability, (ii) the need for a qualitative approach incorporating quantifiable measures to enhance the usefulness of the analysis, (iii) loose semantics of the proposed model, leading to consistency issues when different people use it, and (iv) lack of validation mechanism with stakeholder involvement for evaluating decisions and their impacts. 

\citet{Penzenstadler_2013} present a meta-model that is instantiated as a sustainability model to analyze sustainability dimensions, value indicators, and support activities. The meta-model describes the sustainability dimension as an aspect of the goal. Indicators are used to quantify a value that represents certain sustainability dimensions. These indicators, guided by external regulations, guide the activities that lead to achieving sustainability. It is, however, challenging to identify possible conflicts between goals. To solve this issue, \citet{Saputri_2016} present an extension of this meta-model by including QAs as a measurement mechanism for indicators. A template is provided to carry out this trade-off analysis between goals by capturing the influence (conflict, support, neutral). 

\citet{Duboc_2020} present a framework to raise awareness on the sustainability impacts of software systems in the context of requirements engineering. They present a question-based Sustainability Awareness Framework (SusAF) supplemented by Sustainability Awareness Diagrams. This framework aims to facilitate discussions, for it to become a de-facto standard in the industry as a thinking tool for short, medium, and long-term sustainability impacts of socio-technical systems \cite{SusAF_Vision_2023}. 

More recently, the GAISSA project~\cite{GAISSA} provides architecture‐centric methods for the modeling and development of green AI‐based systems. It focuses on developing a greenability quality model for assessing AI models and systems, providing green-aware architecture-centric methods to data scientists and software engineers, and supporting analysis and decision-making for greenability improvement. 

\citet{GuldnerEtAl_DevelopmentEvaluation_2024a} provide a reference measurement model called the Green Software Measurement Model (GSMM). The model builds upon existing research to leverage recommendations, processes, and tools to help evaluate and minimize software's environmental impact. The study presents a catalog of metrics to evaluate software's energy and resource consumption. Further, it provides a list of tools for automation, resource logging, analysis, and fully integrated solutions. As such, GSMM aims to aid practitioners in developing and evaluating software products for resource and energy efficiency. 

\citet{Fatima2024-sus-cs} perform a software architecture assessment for sustainability through a case study. The study evaluates architecture design decisions to identify the inter-QA trade-offs. It presents a sustainability impact score (SIS) to quantify and compare the QA trade-offs across the four sustainability dimensions. The SIS can aid SA evaluators in making sustainability-aware decision-making.

The frameworks and models found in the literature facilitate capturing and representing software sustainability in different stages of the Software Development Life Cycle. The existing frameworks, however, lack instruments for targeted decision-making regarding the concrete software elements~\cite{Naumann_2011, Cabot_2009, GuldnerEtAl_DevelopmentEvaluation_2024a}. Current research includes thinking frameworks that can aid in decision-making, but with a focus on requirements \cite{Penzenstadler_2013, Duboc_2020, Saputri_2016}.  
We observe a missing link between goals and the software measurement process, with software architecture at the center stage. Establishing a connection between micro-level metrics that represent the sustainability of architectural elements to macro-level organizational sustainability goals, is essential for improvement and taking action. Determining the level of detail is a challenge observed in all frameworks as low-level details tend to get disconnected from high-level views of the systems and organizations.

With the SAF Toolkit, we envision support along certain aspects, specifically in the context of software architecture by (i) measuring the Quality Requirements (QRs) representing design concerns via metrics (already supported by the current release of the SAF Toolkit), (ii) establishing a link of design concerns to software architecture and its elements (partially supported by the SAF Toolkit), (iii) connecting the low-level metrics measuring the QRs with high-level goals represented by KPIs (in the future release of SAF Toolkit). 

\section{The {SAF} Toolkit}\label{sec:toolkit}

To provide a thorough introduction to the SAF Toolkit, we organize this section in three parts. First, we take a bird's eye view of the toolkit to understand all instruments involved and how they interact. Then, we zoom into each instrument individually to examine their elements and show how the instrument has evolved over time to the current release of the toolkit. Finally, we outline our plan for the future and how we envision the planned extensions.

\subsection{The Big Picture}\label{sec:bigpicture}

\begin{figure}[H]
    \centering
    \includegraphics[width=1\linewidth]{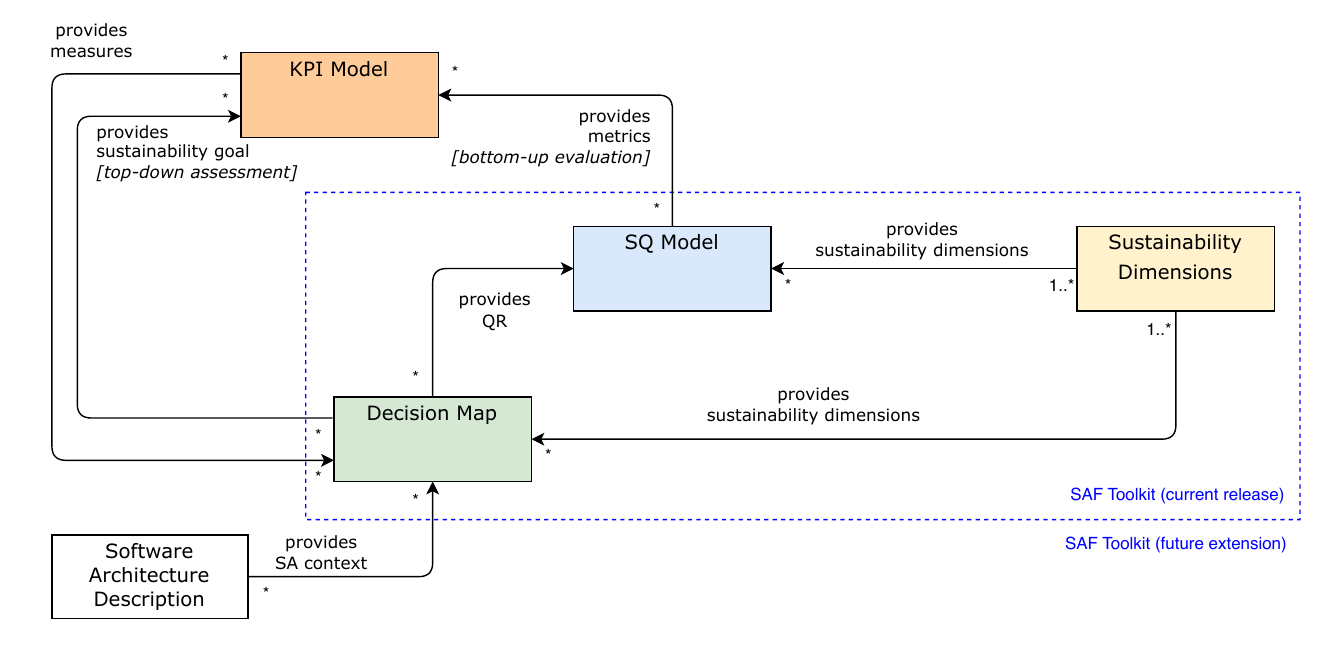}
    \caption{\textbf{SAF Toolkit Overview}. High-level overview of all involved instruments and their relation. Instruments inside the dashed-blue rectangle are available in its current version; instruments outside are plan of our future vision and have yet to be integrated. The figure builds upon the notation for Unified Modeling Language (UML) version 2.5.1 \cite{OMG-UML251}.}
    \label{fig:structuralModel}
\end{figure}

\Cref{fig:structuralModel} provides an overview of all the instruments that are part of the SAF Toolkit and those that are planned as future extensions. We refer to this view as the \textit{SAF Toolkit Overview}, as it presents the instruments at a high level of abstraction and illustrates their inter-relationships.

In its current release (see inside the dashed blue rectangle), the SAF Toolkit consists of its two main instruments, namely the Sustainability-Quality (SQ) Model and the Decision Map (DM), both adopting a shared definition of the sustainability dimensions. This definition is used to classify the QRs from the perspective of the sustainability dimension to which the QRs relate.

In particular, to frame and illustrate sustainability-relevant design and QRs, the visual notation of \textbf{DMs} is utilized. DMs serve as a thinking framework to capture, reason, and uncover dependencies among sustainability-related QRs in the form of a visual diagram. This visualization is also intended to foster communication among software architects regarding the sustainability of their software-intensive systems.

To consolidate definitions about the identified concerns, the \textbf{SQ Model} encompasses a collection of QAs\footnote{In the context of the SAF Toolkit, we use interchangeably terms \textit{quality requirement} and \textit{quality attribute} depending on the context of the discussion. This follows the observation by \citet{boer_vanvliet2009}, that quality requirements become quality attributes during the design phase of a given software project.} grouped into the four sustainability dimensions. QAs are derived in accordance with the ISO/IEC 25010 standard \cite{ISO_25010_2011} and are based on QAs identified in prior projects \cite{Condori-FernandezLago_CharacterizingContribution_2018, Condori-FernandezEtAl_ActionResearch_2020}. The SQ Model operationalizes the identified QAs provided by the DM by assigning a set of metrics and measures.

Research indicates that sustainability necessitates a careful assessment of trade-offs but within distinct \textbf{Sustainability Dimensions} \cite{LagoEtAl_FramingSustainability_2015, BetzEtAl_SustainabilityDebt_2015, Condori-FernandezLago_CharacterizingContribution_2018}. As mentioned in \Cref{sec:Q-dim}, different disciplines feature varying types and numbers of dimensions. The SAF Toolkit, with its DM and SQ Model aligns with the four sustainability dimensions as proposed by \citet{LagoEtAl_FramingSustainability_2015}. Nonetheless, while all four dimensions should be taken into account during the design phase to achieve balance, it is not always necessary for every dimension to be relevant for a specific software system \cite{Condori-FernandezLago_CharacterizingContribution_2018}. 
\Cref{sec:SAFnow} delves deeper into the current release of the SAF Toolkit by further explaining each instrument.

The instruments yet to be integrated (see outside the dashed-blue rectangle in \Cref{fig:structuralModel}) include the KPI Model and the Software Architecture description. To establish and define the metrics and measures for the QAs within the SQ Model in a systematic way, we aim to incorporate a \textbf{KPI Model}. Such a model will assist us in identifying sound KPIs relevant to the specific software project at hand. 
KPIs can emerge from two distinct approaches, top-down or bottom-up. 
In a top-down assessment, we envisage that KPIs can be formulated by starting with a sustainability goal derived from the DM, followed by gathering the relevant metrics and measures. 
Conversely, in a bottom-up assessment, concrete metrics from the SQ Model would be used to formulate, or associate to, a particular goal. 
Overall, once comprehensive KPIs are established in a systematic and traceable manner, they can provide essential feedback on QAs and related design decisions through tangible measures.

To contextualize the software architecture of a given software-intensive system, we aim to enhance the SAF Toolkit by including the \textbf{Architecture Description} that makes explicit the relevant architecture elements realizing the design decisions responsible for sustainability impacts. To provide this integration, we adopt and integrate architecture elements in compliance with the ISO/IEC/IEEE 42010:2022 Standard for Software Architecture Description~\cite{ISO_42010_2022}.
If present, this integration would enable us to link the QAs from the DM with specific architecture elements, such as software services or components. This connection is crucial for (i) understanding where to measure the previously defined KPIs and consequently (ii) determining the impact of a specific architecture element on a particular QA.
The detailed conceptualization and integration for the planned extensions are elaborated upon in \Cref{sec:SAFfuture}.




\subsection{The SAF Toolkit: Where we are now}\label{sec:SAFnow}
The current release of the SAF Toolkit produces two main outputs, the DM and the SQ Model. To do so, the toolkit user is provided with a set of so-called instruments, \ie tools providing the necessary guidance. 
The following explains both outputs and for each of them, describes (i) the conceptual elements they frame, (ii) the visual elements with examples from some of our past projects, and (iii) the instruments guiding their creation. We close with a reflection on the experience and the points for improvement related to each instrument.

\begin{sidewaysfigure}[htbp!]
    \centering
    \includegraphics[width=1\linewidth]{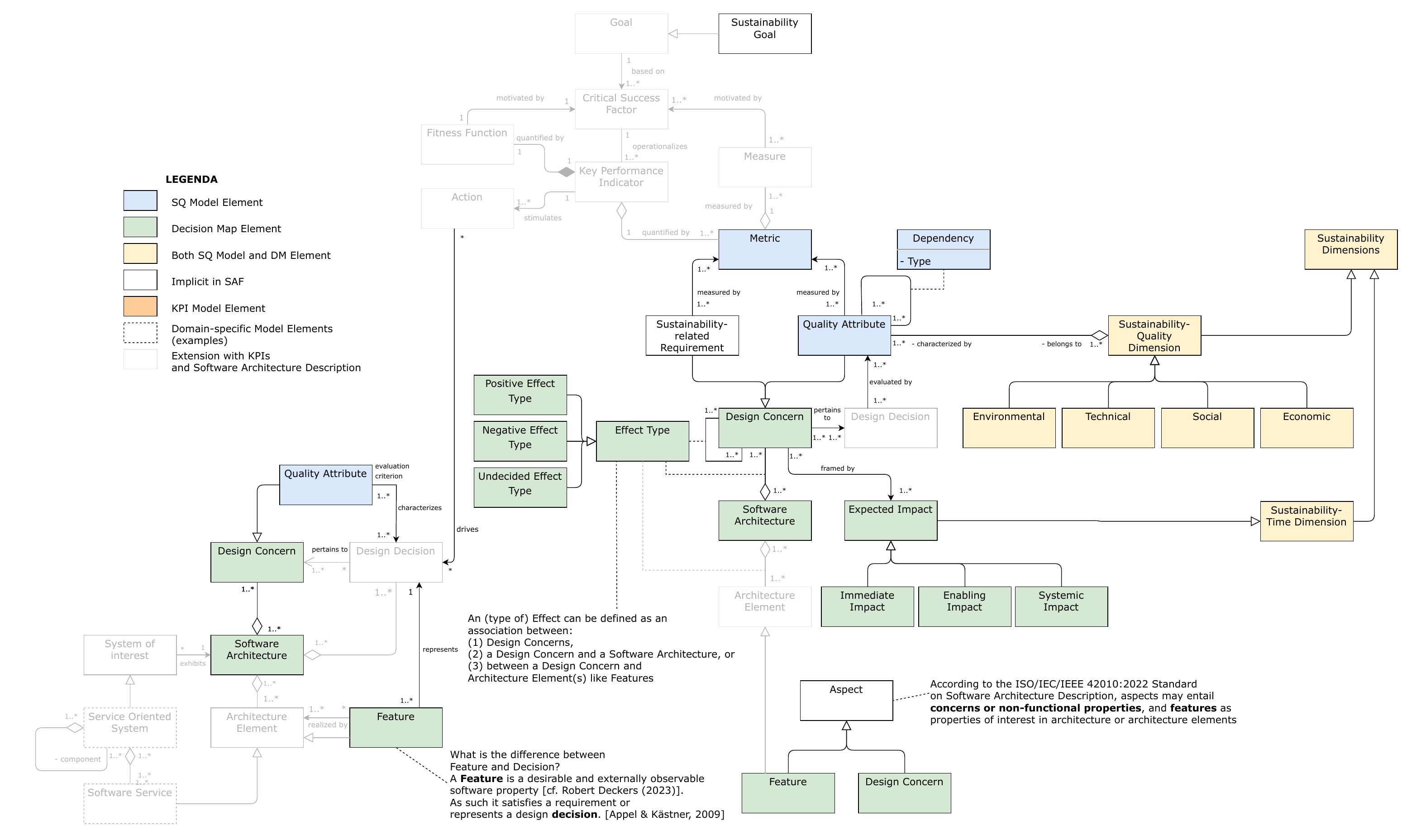}
    \caption{\textbf{SAF Toolkit Model.}} Detailed-level overview of all involved instruments, their elements and relation. Colored elements are available in its current version; Transparent elements are plan of our future vision and have yet to be integrated. The figure uses the notation for Unified Modeling Language (UML) version 2.5.1 \cite{OMG-UML251}. A high-resolution version of this figure is included in the supplementary material online~\cite{saf_toolkit_online}.
    \label{fig:toolkitModel}
\end{sidewaysfigure}

\subsubsection{Decision Map and its Instruments}

\paragraph{The Conceptual Elements}
In \Cref{fig:toolkitModel}, the key elements of the DM (coloured in green) are encompassing architecture-related concepts widely adopted in the field, as well as the essential sustainability-related concepts.
The architecture-related concepts include class \texttt{Software Architecture}\footnote{For presentation clarity, and to facilitate traceability with \Cref{fig:structuralModel}, some classes in \Cref{fig:toolkitModel} are repeated. However, classes with the same name also represent the same class.} as the generic \textit{anchor} to the software project under study (or if you wish, the generic placeholder representing the software architecture), and class \texttt{Design Concern} representing the set of architecture design concerns that are being addressed. In addition, a DM may include elements of class \texttt{Feature} (see the bottom and left-hand side of \Cref{fig:toolkitModel}), which according to the ISO/IEC 25010 standard is an aspect or property (functional or non-functional) realized by architecture elements and as such part of the software architecture. 
In the current release of the SAF Toolkit, a DM can be designed to illustrate \texttt{Design Concerns} that may address both \texttt{Sustainability-related Requirements} (\eg in electric vehicle (EV) charging services, how to ensure that any type of citizen can easily understand how to charge), and \texttt{Quality Attributes} (\eg usability).

The sustainability-related concepts, in turn, are represented by association class \texttt{Effect Type} and class \texttt{Expected Impact}. The \texttt{Effect Type} represents the type of sustainability effect between a pair of design concerns (\eg the effect between performance and energy efficiency). Such effect can be measured (or designed to be) positive (\ie contributing to sustainability), negative (\ie hindering sustainability) or undecided (\ie if no measures are taken yet and at design-time it is unclear what type of effect will be observed).
The \texttt{Expected Impact}, in turn, represents the sustainability time dimension and is meant to frame if the impact of a certain design concern is immediate (first order), enabling (second order) or systemic (third order). 

\begin{figure}[ht!]
    \centering
    \includegraphics[width=1\linewidth]{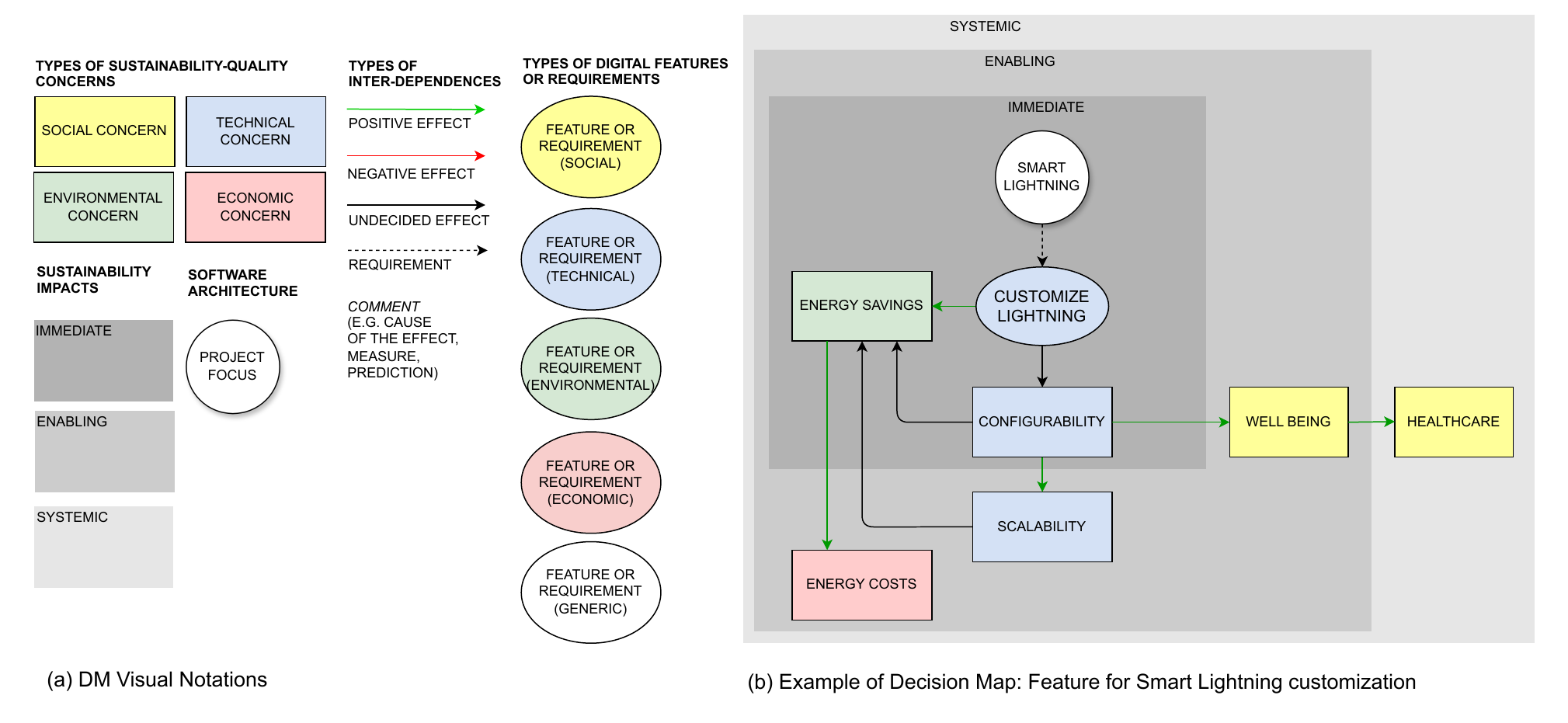}
    \caption{DM visual notation and a simple example}
    \label{fig:DMnotationANDexample}
\end{figure}

\paragraph{The Decision Map}
By focusing on a selected software project, the DM can illustrate the features that should be sustainability aware. As shown in \Cref{fig:DMnotationANDexample}, a DM makes use of a relatively simple visual notation (summarized in \Cref{fig:DMnotationANDexample}.(a)) so that any type of stakeholder, from IT specialists to general decision makers, can use it to reason (i) about the sustainability implications of their design decisions, and (ii) about the decisions' effect on QAs in the short- and longer term.

As an example, \Cref{fig:DMnotationANDexample}.(b) shows the DM for a smart lighting software architecture project, and in particular the feature \texttt{customize lightning} and the related network of sustainability effects that such feature is expected to have on various relevant QAs. For example, the feature is expected to have a \textit{positive effect} on energy savings in the environmental Sustainability Dimension (\ie one can avoid electricity waste by tuning lightning to the actual user needs), which in turn would have a \textit{positive effect} on energy costs in the economic Sustainability Dimension (\ie by saving energy one would reduce energy costs, too).

Further, QAs are mapped on the Time Dimension to enable explicit representation of the type of sustainability impacts that they are designed for. In our example in \Cref{fig:DMnotationANDexample}.(b), configurability of lightning to human behavior has proven to enable positive (second order) effects on well being (\eg on sleeping and eating patterns of children) which in turn may have systemic (third order) effects on the need for (and cost of) healthcare.

\paragraph{The guiding Instruments}\label{sec:guidingInstrument}

The SAF Toolkit includes a number of instruments that provide guidance for creating a DM. These instruments are the Checklist and the Decision Graph, both described in the following.

\begin{description}
    \item \textbf{The Checklist} provides a list of reflective questions that help guiding the creation of a DM in a systematic way. It mainly focuses on the identification of the most important QAs and design concerns. \Cref{fig:DM-instruments}.(a) shows a fragment of the Checklist with the reflective questions for identifying the main QAs (and the types of related sustainability effects) for the software project being considered.

\begin{sidewaysfigure}[htbp!]
    \centering
    \includegraphics[width=\linewidth]{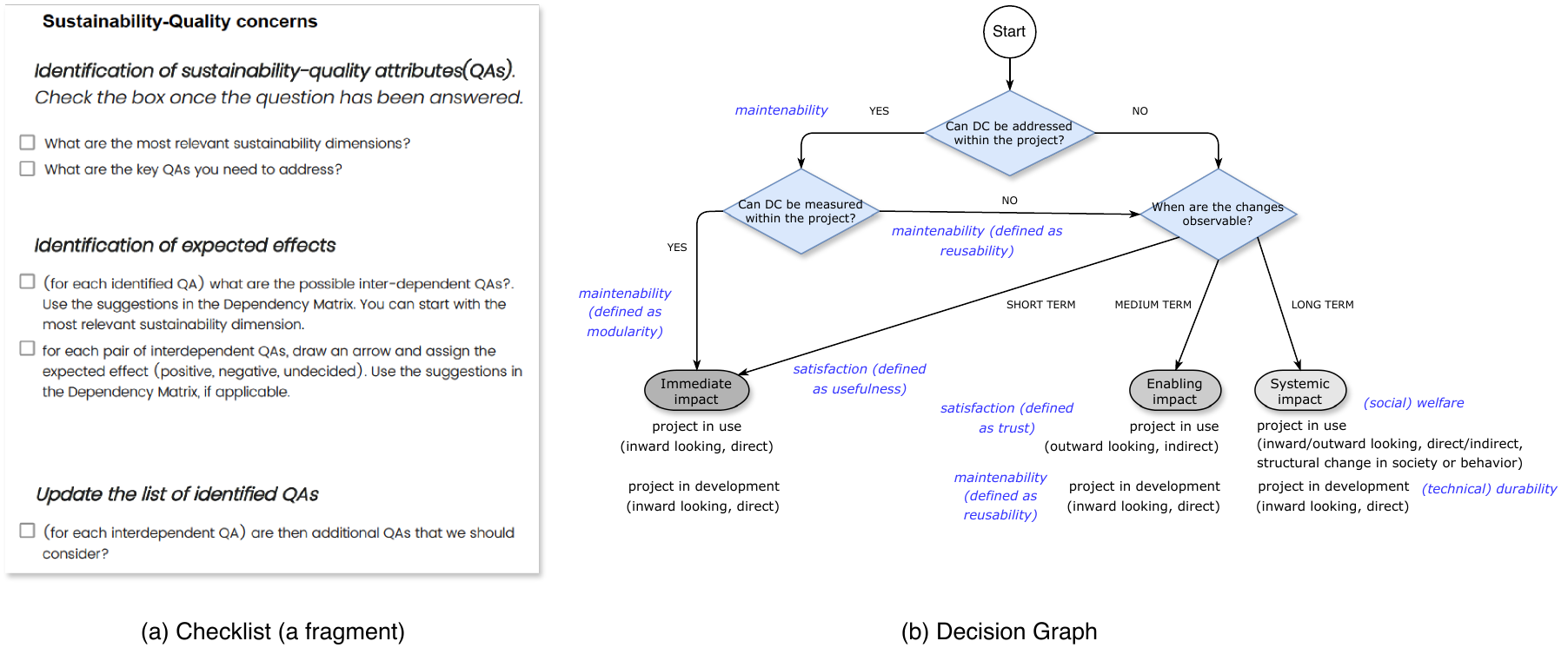}
    \caption{\textbf{DM instruments.} (a) Partial view of a checklist used in the DM creation process. (b) Decision graph (flowchart) to determine the type of sustainability impact. The figure uses a flowchart notation (cf. Wikipedia) where diamonds represent decision points (questions), arrows indicate the responses and flow direction, and rounded rectangles represent the impact levels.}
    \label{fig:DM-instruments}
\end{sidewaysfigure}

    \item \textbf{The Decision Graph} provides guidance on classifying each QA according to the most relevant Time Dimension. Practically, it helps deciding where each selected design concern should be placed among the three impact levels (\ie immediate or first-order, enabling or second-order, systemic or third-order -- the decision graph leaves in \Cref{fig:DM-instruments}.(b)). \\
    In particular, the black text in the figure exemplifies the possible interpretations of each impact level; while the blue text provides a simple example of possible mappings. As such, if the QA would be \texttt{maintainability}, depending on its definition, its impact could be measured directly in the software system (\eg if defined as the level of modularity of the design), indirectly (\eg if defined as reusability and hence measurable only after the system is reused in new projects), or systemic (\eg if defined as durability and hence measurable in terms of the stability~\cite{Salama2021} of the deployed system over longer time periods).
\end{description}

\begin{figure*}[ht] 
  \begin{subfigure}[b]{0.45\linewidth}
    \centering
    \includegraphics[width=0.98\linewidth]{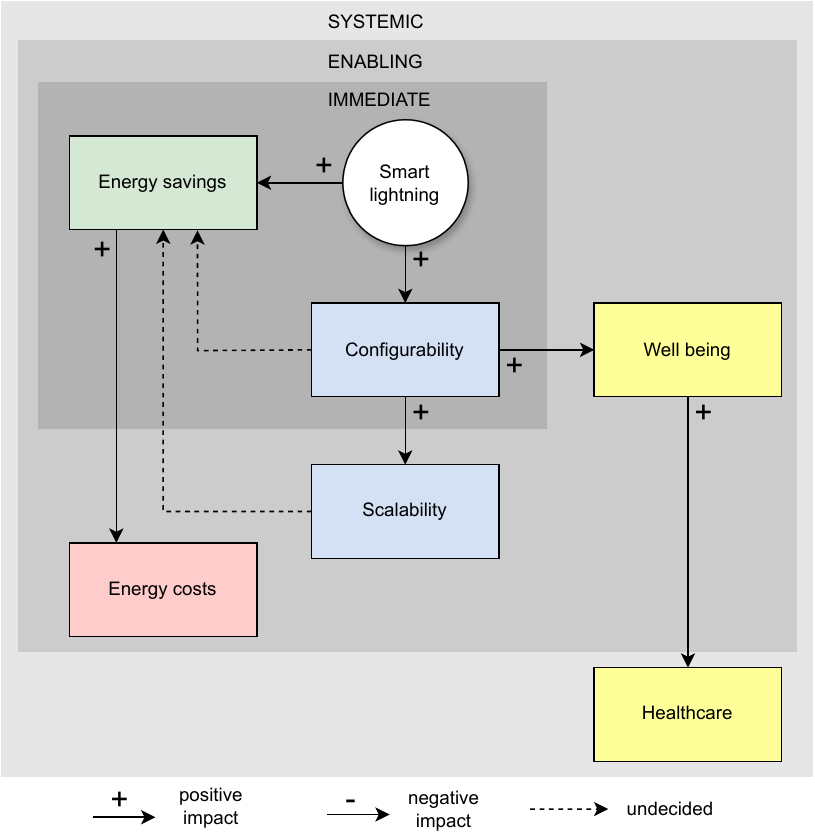} 
    \caption{DM: only Design Concerns} 
    \label{fig:DM.DC} 
    \vspace{4ex}
  \end{subfigure}
  \begin{subfigure}[b]{0.55\linewidth}
    \centering
    \includegraphics[width=0.98\linewidth]{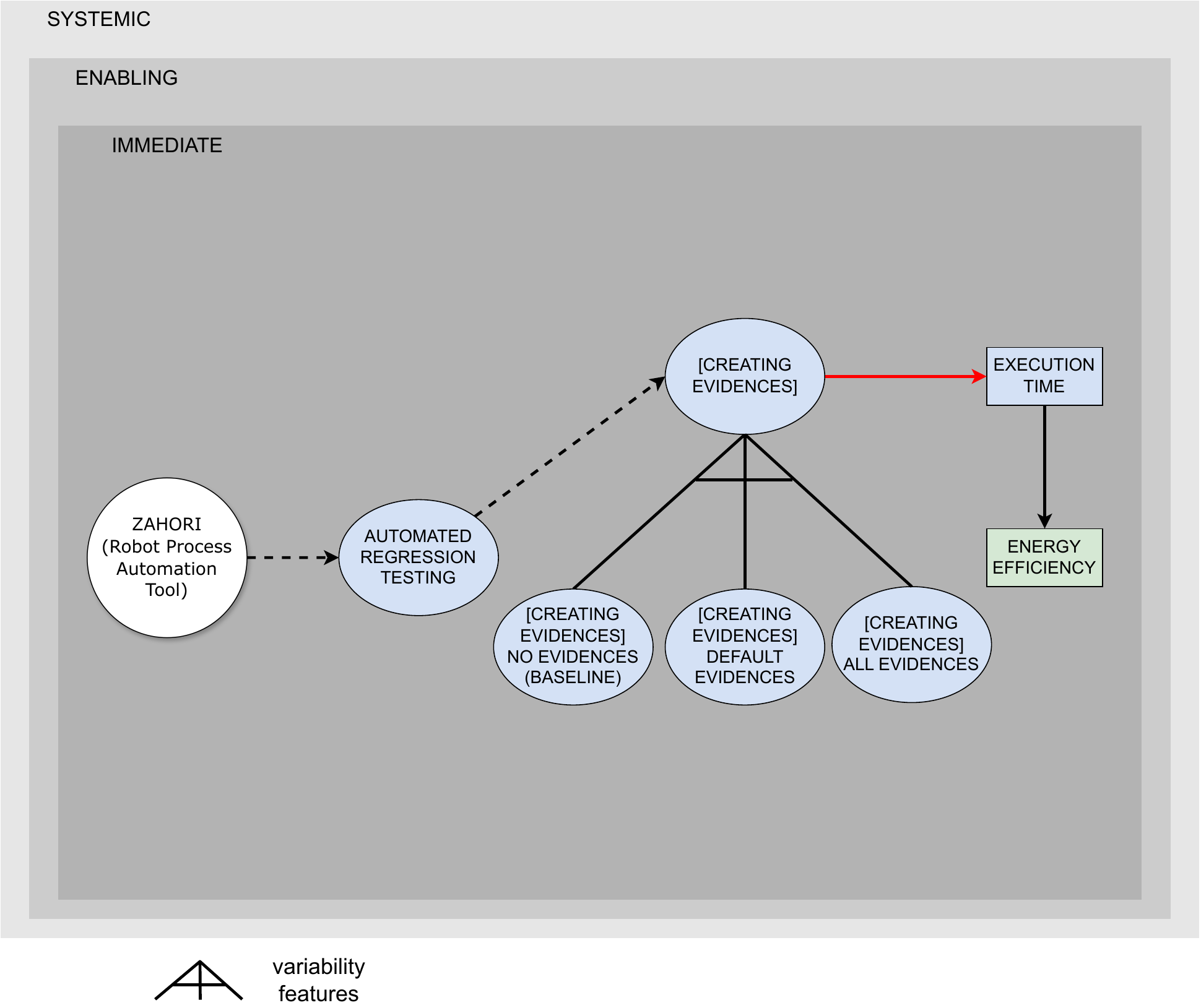} 
    \caption{DM: extension with variability feature} 
    \label{fig:DM.var} 
    \vspace{4ex}
  \end{subfigure}
\vspace{-30pt}
  \caption{Decision Maps. [See notation in Fig.~\ref{fig:DMnotationANDexample}.(a).]}
  \label{fig:DM.before} 
\end{figure*}

\paragraph{Reflections}
It must be noted that the DM notation has evolved over the years to address the needs of sustainability-aware software architecture design decision making, thanks to the experience gained via its use in a series of diverse cases provided by external partners, mostly from industry. For example, while the toolkit was created starting in 2015, only in 2020 we introduced the notion of `feature' to make explicit which aspect of a software system would be responsible for which QAs. Accordingly, before 2020 the DM in \Cref{fig:DMnotationANDexample}.(b) was not including any feature and was looking like the one in \Cref{fig:DM.before}.(a)\footnote{Notice that before 2020 also the visual notation for the effect types was different; later on improved thanks to the feedback received from the practitioners involved in the case studies.}. 
As another example, the more recent work reported in \citet{FunkeEtAl_VariabilityFeatures_2023} had to represent the fact that some features can be realized by multiple variants, as illustrated in \Cref{fig:DM.before}.(b).

In \Cref{tab:cases} we summarize a list of exemplary cases from the case studies (classified as such by using the checklist proposed by \citet{Wohlin2022-CS-checklist}) that we have carried out over the years to apply the SAF Toolkit and, when needed, to improve it according to the lessons we learned. On the right-hand side of the table, we report if and how the SAF Toolkit was revised to accommodate needs that we deserved as generally applicable and adding value. In particular, changes to the DM have been: 
\begin{itemize}
    \item In 2017, a case study carried out with the KPMG consultancy company revealed that after carrying out empirical experiments, we could decorate the effect-arrows with labels that would \textbf{illustrate the measured impact} \eg of a quality concern like performance on another quality concern like energy efficiency. As a consequence, we added impact labels to the DM visual notation (see \texttt{comment} in \Cref{fig:DMnotationANDexample}.(a)).
    \item In 2020, we conducted a research project on food security, in particular in resource-constrained environments (like some African countries), where access to software services can help sustainability and equity. We had to model which \textbf{features} were meant to address equity requirements (\eg provide farmers with audio-based access to services in various local dialects) so that all farmers could exploit the opportunities of selling online their products. It must be noted that in the same project, we experimented with modeling with goals in DMs, too, \ie making explicit the \textbf{sustainability goals} that features are meant to realize. While this modeling exercise was beneficial for this specific project, we did not encounter such a modeling need afterwards, hence we did not leverage the notion of sustainability goal in the toolkit yet. This decision is driven by our commitment to maintaining the simplicity of DM notations, ensuring they remain understandable for both technical and non-technical stakeholders.
    \item In some other projects between 2020 and 2022 we found it useful to, \eg model design concerns that are in a hierarchical relation (like that existing between quality characteristics and sub-characteristics) or in the form of \textbf{thematic clusters} (like complexity into organisational, technical or social complexity). While such representations could be model kinds of DMs, we consider them as different views. Also in these cases, we did not need to illustrate them in any other project and hence refrained to extend the DM notation.
    \item In 2020-2023 we carried out a yearly winter school attended by both researchers and practitioners. The participants would bring their own case study. Every school edition, we had one case where modeling the DM elements pertaining a \textbf{stakeholder type} would be beneficial (\eg if some quality concerns would pertain producers and some others consumers). While also in this case we refrained from leveraging the \textbf{clustering of concerns} in the standard DM notation, we noted down this experience as a lesson potentially useful for future releases of the toolkit. 
\end{itemize}

\begin{table}[htbp!]  
\centering 
\begin{adjustbox}{width=1\textwidth}
\small
\begin{tabular}{l p{2.5cm} p{5cm} cc c | cc} 
\hline\hline 
\multirow{2}{*}{YEAR} & \multirow{2}{*}{\textbf{TOPIC}} & \multirow{2}{*}{\textbf{SHORT DESCRIPTION}} & \multicolumn{2}{c}{\textbf{CASE ORIGIN}} & \multirow{2}{*}{\textbf{REFS.}} & \multicolumn{2}{c}{\textbf{SAF TOOLKIT}}
\\  
& &  & Industry & Academia & & DM & SQ Model 
\\
\hline   
2016 & SoSA exercise & Tutorial at International Conference on Advanced Information Systems Engineering (CAiSE) 2016, where three working groups used a smart mobility case to critically evaluate DMs. &  & \checkmark & \cite{Sergio_Espana2016} & \checkmark &  \\\rowcolor{gray10}
2017 & KPMG Qubus & Empirical experiment assessing the impact on energy consumption and performance of different releases and deployment strategies of an industrial software product  supporting process management. & \checkmark &  & \cite{Verdecchia2017} & \checkmark +I labels &  \\
2017 & Sustainability-aware Mobility & Case study from the City of Amsterdam, exploring mobility as a service provided to the citizen to improve the safety, accessibility, air quality, quality of life, and attractiveness of Amsterdam. & \checkmark & & \cite{Niggebrugge2018} & \checkmark &  \\\rowcolor{gray10}
2018 & Smart Work & Scenarios to explore the implications of virtual mobility, \ie substituting activities that require physical mobility with others that allow for virtual presence \cite{Lago2019} & & \checkmark & \cite{Lago2019} & \checkmark & \\
2020 & Food Security & Case study to identify the sustainability and ethical concerns of designing context-sensitive ICT systems in low-resource environments. The general goal is to define a general good software design practice. & \checkmark &  & \cite{Vos2020-tr,Vos2020-paper} & \checkmark +F,G & \\\rowcolor{gray10}
2020 & Responsible Flight Planning & The SAF Toolkit was applied in two workshops with experts from a multinational company specialized in operations- and flight planning products and software in the aviation sector. The goal was to identify sustainability opportunities and challenges for their flight planning software. & \checkmark &  & - & \checkmark & \checkmark \\ 
2021 & Information System (IS) Complexity and Sustainability & Based on the experience of IS professionals, this project identified a reusable taxonomy of concerns that are found to influence IS complexity and sustainability. & \checkmark &  & \cite{Bischoff2021} & \checkmark +C hierarchy & \checkmark \\\rowcolor{gray10}
2020-2022 & VU Winter Schools & Application of the SAF Toolkit to: four industrial cases in the domains energy provisioning, tourism, IT services, SDG reporting (2020); seven cases spanning from energy-efficient search engine, trustworthy open digital news, usable and sustainable opensource software platforms, to city-wide car sharing (2021); and four cases from paperless purchase receipts to digital mobile currency (2022). & \checkmark & \checkmark & \cite{Lago2021-winterschool} & \checkmark +C clustering & \checkmark \\ 
2020-2023 & Student Projects: Digitalization and Sustainability & 67 student group projects over four Master-level course editions, focused on digitally transforming aspects of various domains, from agriculture, education, retail, to entertainment, music festivals, arts. &  & \checkmark & - & \checkmark +T & \checkmark \\\rowcolor{gray10} 
2023 & Variability Features & In an industrial case study the notion of variability features and variants (\ie software features that can be implemented in different alternatives \cite{FunkeEtAl_VariabilityFeatures_2023}) were introduced by applying and extending DMs. The SQ model was used to establish a measurement plan and monitor the impact of each emerged variant over time. & \checkmark &  & \cite{FunkeEtAl_VariabilityFeatures_2023} & \checkmark & \checkmark \\
\hline\hline 
\end{tabular}  
\end{adjustbox}
\caption{Case studies experimenting with the SAF Toolkit. Toolkit extensions (indicated with prefix `\textbf{+}') entail \textbf{I}=Impacts, \textbf{F}=Features, \textbf{G}=Goals, \textbf{C}=Concerns, \textbf{T}=Technologies} 
\label{tab:cases}
\end{table}  

\subsubsection{Sustainability-Quality Model and its Instruments}

\paragraph{The Conceptual Elements}
In \Cref{fig:toolkitModel}, the key elements of the SQ Model (coloured in blue) are encompassing QAs, dependencies among them, and associated metrics. QAs, in turn, are characterized according to the sustainability dimension they belong to.

\paragraph{The SQ Model}
The SQ Model comprises a collection of \texttt{QAs} categorized into four \texttt{Sustainability-Quality Dimensions} (\eg technical dimension includes security, environmental dimension includes energy efficiency). Recognizing the potential interdependencies among QAs and the multidimensional nature of software sustainability, such dependencies (\ie \texttt{Dependency} instances) can be of two types:

(i) \textit{Inter-dimensional dependencies}, linking a pair of QAs defined across different dimensions (\eg robustness from the technical dimension positively influencing confidentiality in the social dimension).

(ii) \textit{Intra-dimensional dependencies}, where a dependency emerges between two distinct QAs within the same dimension (\eg within the technical dimension, security may depend on reliability).

These distinctions are relevant for two reasons: (i) understanding inter-dimensional dependencies helps in comprehending how improvements in one dimension can positively impact another, promoting holistic sustainability in software architecture; (ii) recognizing intra-dimensional dependencies allows for focused improvements within specific dimensions, enhancing the overall quality and sustainability of software systems.
Moreover, as our SQ model provides support to both identify design concerns and assess QAs of the software architecture, a set of \texttt{Metrics} is used for measuring QAs. Similarly, \texttt{Metrics} can be defined for \texttt{Sustainability-related requirements}, represented as a class inheriting from class \texttt{Design Concern}.

\paragraph{The guiding Instruments}
The SAF Toolkit further includes a number of instruments that provide guidance for the software architect to produce an SQ Model. These are the SQ Model Template and the Interdimensional Dependency Matrix Template, both described in the following.

\begin{description}
    \item \textbf{The SQ Model Template} provides a structured approach to assist software architects in creating their own specific SQ Model. It comprises three well-defined components:
    \begin{itemize}
        \item \textit{Quality Attributes:} A curated list of QAs considered during the design or assessment of sustainability-aware software systems. Each attribute is accompanied by operational definitions delineating its intended purpose.

        \item \textit{Direct Relationships to Sustainability Dimensions:} Explicit mappings between QAs and overarching sustainability dimensions such as economic, environmental, social, and technical.

        \item \textit{Metrics:} Defined to offer a scale and method for measurement, the metrics encompass three types according to the standard ISO/IEC 25010:2011: internal metrics, external metrics, and quality-in-use metrics.
    
    \end{itemize}

    Overall, the template (see \Cref{fig:SQmodel_Zahori} for an example from \cite{FunkeEtAl_VariabilityFeatures_2023}) is structured in a tabular format and consists of five fields: the name of the QA, its definition, the reference to the source of definition, the corresponding sustainability dimension, and a list of metrics.

    Software architects have the flexibility to customize their SQ Model by either starting from a generic SQ Model that consolidates QA definitions from past projects and ISO/IEC 25010-2011 standards and/or by introducing new QAs not present in the generic model. They can also refine existing definitions and establish corresponding direct relationships.
    Figure \ref{fig:SQmodel_Zahori} lists two QAs: Execution Time
    (ET) and Energy Efficiency (EE) that contribute to technical and environmental dimensions, respectively. Both QAs  were identified as relevant of a specific sub-feature named \textit{variability} in the context of one of our case studies where we analysed a process automation tool named Zahori \cite{FunkeEtAl_VariabilityFeatures_2023}. The metrics for ET and EE are variant specific, both are external metrics.
    
    The SQ Model template transcends a mere static list of QAs related to sustainability dimensions; it serves as a dynamic toolset for informed decision-making and planning for the future. Through the construction of their specific SQ Models, architects can:
    \begin{itemize}
        \item Identify QAs that are critical to achieving sustainability within specific contexts.
        \item Incorporate sustainability considerations proactively into the software design process, hence fostering innovation and resilience in the face of evolving environmental, economic and social challenges.
    \end{itemize}

\begin{figure}[H]
    \centering
    \includegraphics[width=\linewidth]{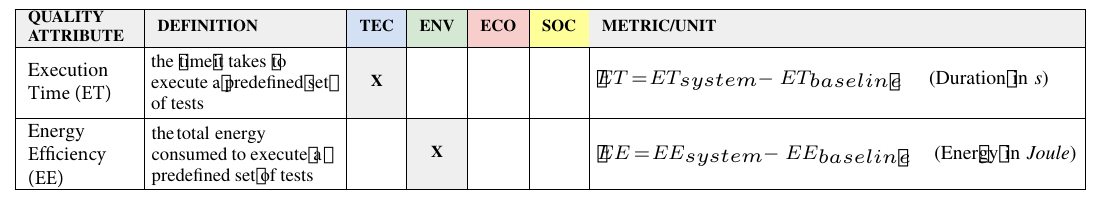}
    \caption{\textbf{The SQ Model} (example revised from \cite{FunkeEtAl_VariabilityFeatures_2023}). The SQ Model template consists of the table structure (headers). The inserted examples are ET and EE (rows). \\Dimensions: TEC=Technical; ENV=Environmental; ECO=Economic; SOC=Social}
    \label{fig:SQmodel_Zahori}
\end{figure}
    
 \item \textbf{The Interdimensional Dependency Matrix (DMatrix) Template}\label{sec:DMatrix} constitutes another vital instrument within the SAF toolkit, designed to facilitate software architects in analyzing and managing dependencies among QAs of the SQ Model. 
 The template allows architects to map out dependencies in a clear and organized manner, including instructions for its use and extension. 
 The DMatrix is organized into rows and columns, each representing a specific QA within the sustainability dimensions D1 and D2, respectively. Three matrices, including \texttt{technical-social, technical-economic, and technical-environmental} were created and evaluated through a nichesourcing study, details  can be found in \cite{CondoriFernandez2024-interdepend}\footnote{In our original study \cite{CondoriFernandez2024-interdepend} we also researched the interdimensional dependencies between sustainability dimensions other than technical (\ie environmental-economic, environmental-social and economic-social). We observed, however, that from the perspective of software architecture design (hence primarily a technical perspective), it made sense to prioritize gaining deeper understanding of the dependencies of the technical dimension with the other dimensions and postpone the focus on developing further the other matrices when such understanding was more mature. From a practical perspective, this research strategy turned out to be useful and practical during the case studies, where in fact, we have not received any particular negative feedback so far. From a research perspective, however, the other interdimensional dependencies are equally important, and subject to ongoing and future work (see also our reflection at the end of this section). }. 
An example of the DMatrix template is shown in~\Cref{fig:DMatrix_TECH-ENV}. The content of each cell of the DMatrix indicates the type of the dependency between the corresponding QAs:
 \begin{itemize}
     \item \textit{Positive Dependency:} A positive dependency between two QAs suggests that improvements or changes in one QA are likely to have positive effects on the other QA. For example, enhancing efficiency, that belongs to the technical dimension,  may positively contribute to another QA of the social sustainability dimension like trust. A positive dependency is represented with the symbol "+".
     \item \textit{Negative Dependency:} Conversely, a negative dependency indicates that changes or improvements in one QA may have adverse effects on the other QA. For example, increasing interoperability of the technical dimension may negatively affect another QA of the environmental dimension such as modifiability. A negative dependency is represented with the symbol "-".
     \item \textit{Indeterminate Dependency:} Some cells may contain the value "I" indicating an indeterminate dependency between pairs of QAs from sustainability dimensions D1 and D2, requiring further contextual information to determine whether the dependency is positive or negative.
 \end{itemize}

\begin{figure}[H]
    \centering
    \includegraphics[width=0.85\linewidth ]{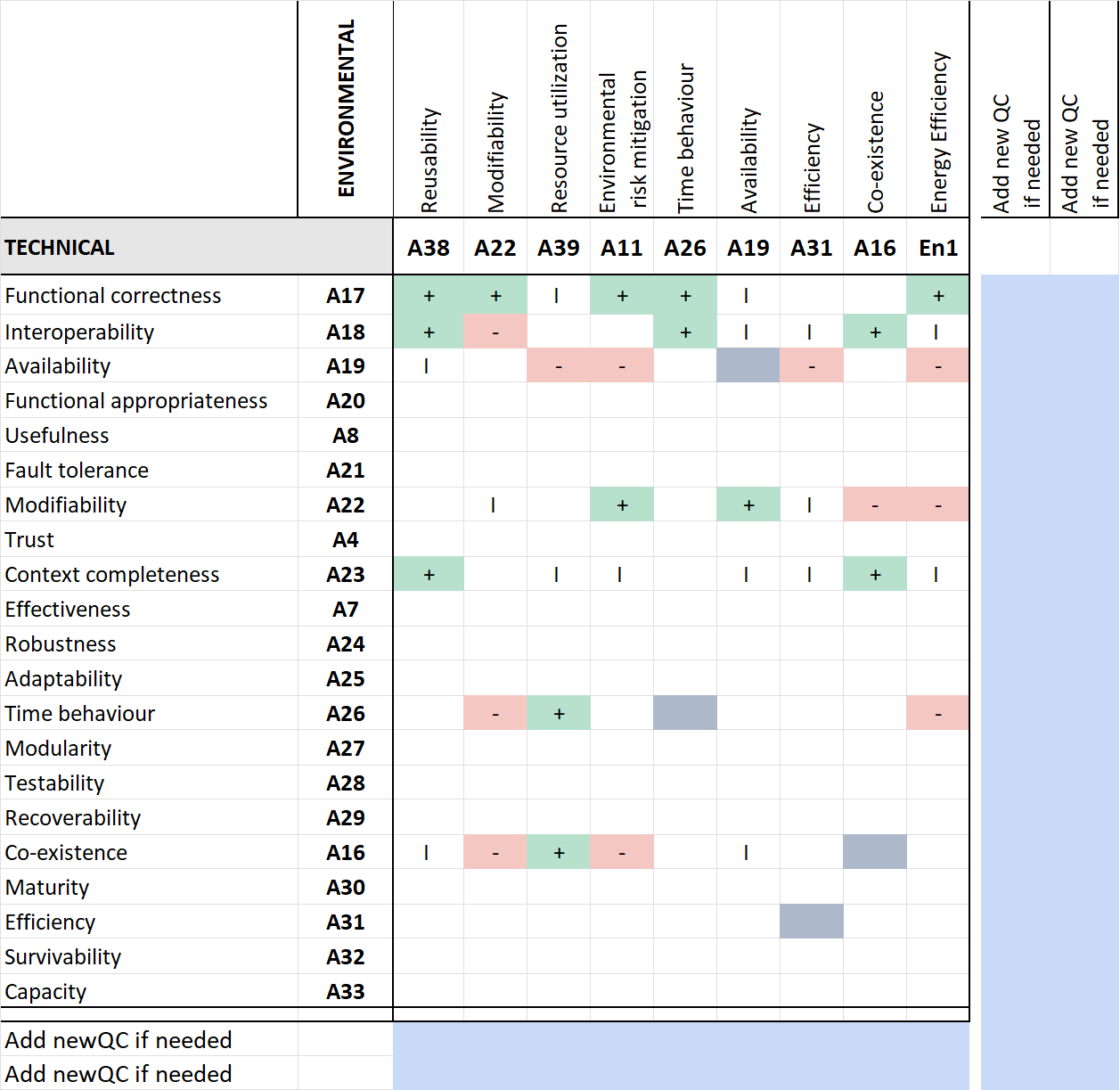}
    \caption{An example of Interdimensional Dependency Matrix template showing the  dependencies among QAs of the technical and environmental dimensions. Cells coloured in blue can be used to add dependencies for newly identified QAs (\cite{CondoriFernandez2024-interdepend}).}
    \label{fig:DMatrix_TECH-ENV}
\end{figure}

By examining the content of each cell, software architects can gain insights into the relationships between different QAs across sustainability dimensions. This understanding helps in decision-making processes and identifying potential trade-offs between sustainability concerns.
 
The DMatrix is instrumental in creating \texttt{DMs}. The nature of dependencies identified in the DMatrix, whether positive, negative, or indeterminate, informs the type of effect between SQ concerns relevant for a project during the DM creation process. 
The DM instruments, i.e., the checklist and decision graph (see \Cref{fig:DM-instruments}) guide the toolkit user through the creation process of the DM and applying the DMatrix accordingly. In \Cref{sec:SAFinAction} we bring all the instruments together and show how they are applied in practice.
\end{description}

\paragraph{Reflections} 
The development and application of the SQ Model template and the DMatrix template have underscored the significance of integrating sustainability considerations into software architecture design. By integrating sustainability dimensions alongside QAs, software architects can make informed decisions to promote technical, environmental, social, and economic sustainability in software-intensive systems. Moreover, and as mentioned before, while we have introduced three matrices focusing on the technical dimension (\ie technical-social, technical-environmental, technical-economic), validation efforts are currently in progress to explore dependencies among QAs across all four sustainability dimensions of SAF. This validation process is becoming increasingly feasible as software companies recognize the significance of addressing sustainability across all dimensions, rather than solely focusing on technical or environmental concerns.

However, there are areas that still demand further attention and development. One such area is the \textit{\textbf{refinement of metrics and measurement techniques}} for evaluating sustainability aspects within software architecture. While existing standards provide a foundation, there is a need for more specific and tailored metrics to capture the unique challenges and opportunities of sustainable software development. 
Another area of improvement is related to the \textit{\textbf{visualization of the SQ Model}}. For instance, graph visualization could offer a more intuitive representation of the relationships among QAs from different sustainability dimensions, facilitating easier analysis and decision-making. Initial exploration in this direction has been presented in \cite{CIBSE22}.
Moreover, it is important to consider that graph-based models enable the persistence of software project data, supporting storage, retrieval, and manipulation for tracking changes over time. This lays the groundwork for future integration of artificial intelligence (AI) techniques, which can leverage graph data to identify patterns, such as common dependencies or critical QRs. 

\subsection{The SAF Toolkit: Where we are going}\label{sec:SAFfuture}

As mentioned in~\Cref{sec:bigpicture}, we are working on two main extensions of the SAF Toolkit which we discuss in more details in Sections~\ref{sec:SAF+KPI} and \ref{sec:SAF+SA}. In addition, we explain how we plan to close the loop between architecture design and re-design, and actual measurement and monitoring of the corresponding system implementation, both driven by measures collected via the integration of the Green Lab (in \Cref{sec:greenlab}) and dashboards (in \Cref{sec:dashboard}).

\subsubsection{Integration of KPIs} \label{sec:SAF+KPI}

\begin{figure}[H]
    \centering
    \includegraphics[width=1\linewidth]{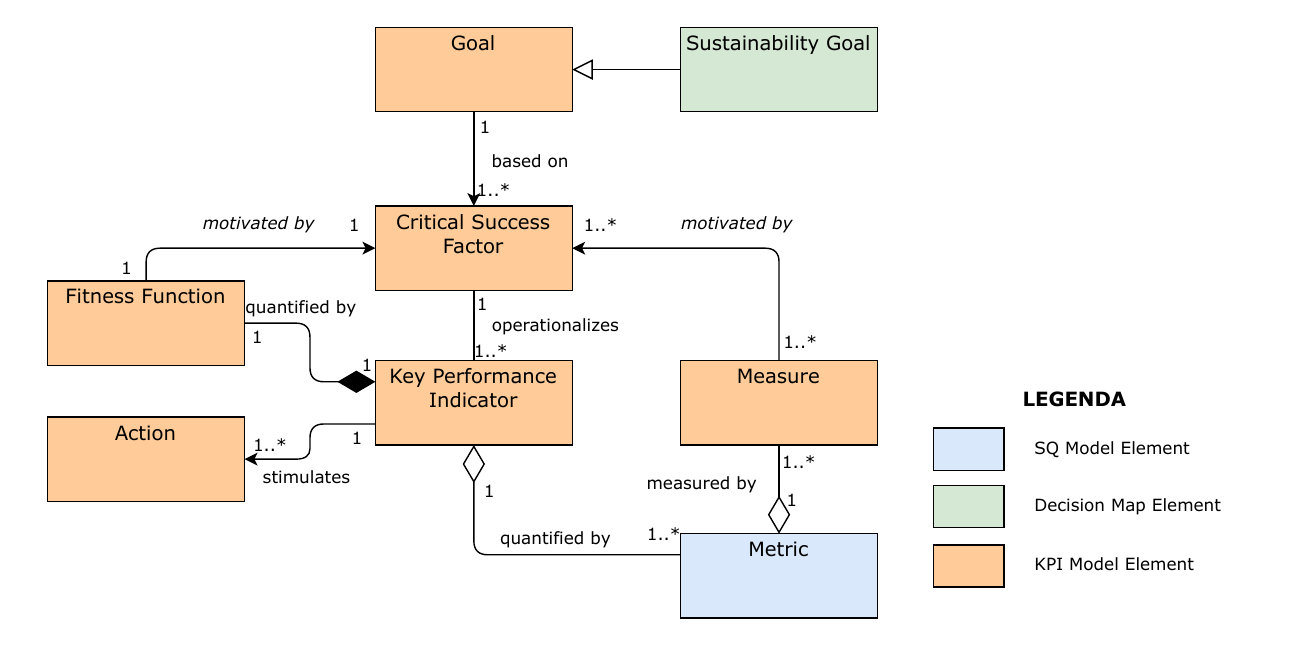}
    \caption{\textbf{KPI Model}. Detailed overview of the envisioned KPI model and its integration into the SAF Toolkit. The figure uses the notation for Unified Modeling Language (UML) version 2.5.1 \cite{OMG-UML251}.}
    \label{fig:KPImodel}
\end{figure}

KPIs are widely used in business organizations to gauge their performance. The integration of the KPI framework \cite{Fatima_2023} with the SAF Toolkit helps embody the notion of sustainability as a software quality in the operational structure of organizations. \Cref{fig:KPImodel} shows the KPI Model, its elements and their relationships with the elements of the SAF Toolkit \ie Sustainability Goal (from DM) and Metric (from SQ Model). 

Business organizations have strategic high-level goals based on the organization's vision and business model. These goals are based on Critical Success Factors (CSFs). 
A CSF defines the critical ingredients that are essential to concretizing and achieving the organizational goal. Achievement of CSFs assumes achievement of success for an organizational goal. This CSF operationalizes the KPI definition and helps motivate the choice of \textit{`what to measure?'} 
These measures can guide the quantification of metrics. 
These metrics can be used to measure the QRs in the SQ Model.
With the integration of the KPI framework within the SAF Toolkit, the same metrics can be used to quantify the KPIs representing one or more QRs. Hence, the KPI is representative of the QRs as a function of the chosen metric(s).

In a daily workflow, the KPI model can be used to formulate context dependent KPIs to monitor the sustainability goals derived by the DM. Once formulated, fitness functions quantify the KPI and check target achievement. 
If targets are not met, the KPI model provides feedback about the modeled design concerns. This feedback triggers appropriate actions (e.g., optimise efficiency of a certain component) in order to influence the selection and implementation of such design decisions.
This helps organizations create a synergy between their sustainability goals and concrete architecture design decisions through an action-driven approach. Supplemented by actions, organizations can take practical steps to improve the software architecture by integrating sustainability as a quality in a goal-focused and design-centric approach. This semantic structure aids in intentional, reusable, and value-focused data collection. 


\paragraph{Reflections}

A KPI is most useful when it is realized as a single value indicator with a threshold-based target. In this way, a single number can represent the achievement of success. This, however, is an optimistic assumption that has some practical obstacles, requiring a nuanced understanding of the system.

\textit{\textbf{How to model the KPI functions?}}
The biggest challenge with the KPI framework is choosing and developing the function to quantify the KPI by aggregating the chosen metrics. There is a need for knowledge-based evidence for the development of KPIs and their fitness functions. The development of KPI functions that are value-focused and relevant requires expert knowledge, regulatory support, and mathematical modeling techniques. 

\textit{\textbf{How to evaluate KPI inter-dependency?}}
While measures are inherently mono-dimensional, metrics can be multi-dimensional. As KPIs are a function of metrics, they become intrinsically multi-dimensional. A KPI definition cannot be isolated for one sustainability dimension. Hence, a tool like DMatrix is not capable of representing KPI interdependency as it only considers two sustainability dimensions at a time. Empirical studies are needed to classify the KPIs and identify their interdependencies. Due to the multi-dimensional nature of KPIs, in the future, we may need different types of dependency matrices to represent inter-KPI dependencies. A systems thinking \cite{checkland1999systems} approach is imperative for the analysis of KPI inter-dependencies and impact analysis of KPIs across sustainability dimensions.

\subsubsection{Integration of Software Architecture Descriptions} \label{sec:SAF+SA}

\begin{figure}[H]
    \centering
    \includegraphics[width=1\linewidth]{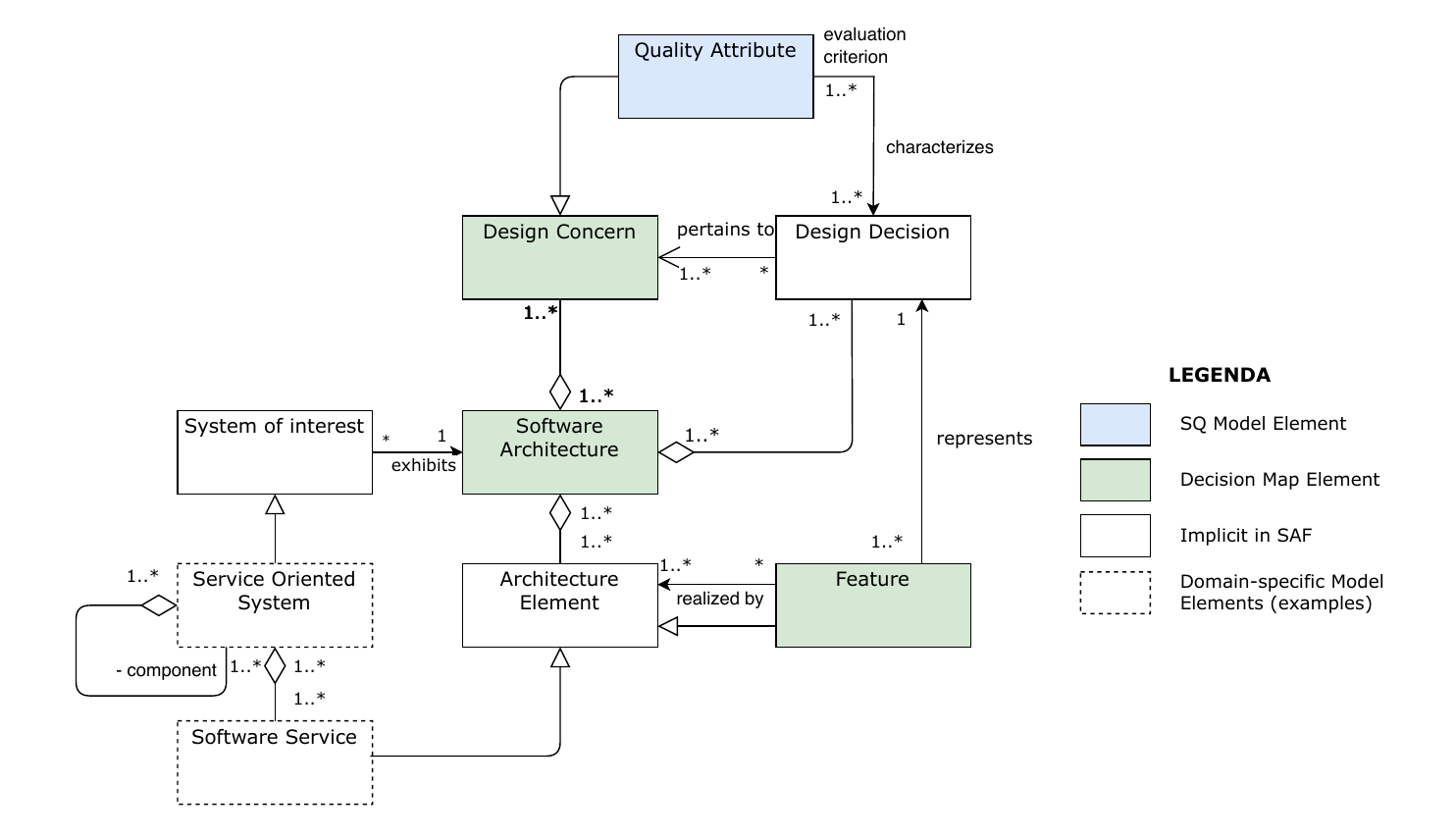}
    \caption{\textbf{Software Architecture}. Detailed overview of the Software Architecture ISO/IEC/IEEE 42010 Standard and its integration into the SAF Toolkit. The figure builds upon the notation for Unified Modeling Language (UML) version 2.5.1 \cite{OMG-UML251}.}
    \label{fig:42010model}
\end{figure}

According to the ISO/IEC/IEEE 42010:2022 Standard~\cite{ISO_42010_2022}, a software architecture consists of multiple design decisions as illustrated in~\Cref{fig:42010model}. The SAF Toolkit incorporates this concept into its DM by integrating design concerns with design decisions. The DM aims to model sustainability concerns during the design phase, with these concerns manifesting as actual decisions in the software architecture at a later stage.

Our toolkit includes the concept of features, which must be distinguished from architectural \texttt{design decisions}. A feature is a desirable and externally observable software property \cite{deckers}. As such it satisfies a requirement or represents a design decision~\cite{Apel2009}. Modeling these features allows architects to (re-)assess their decisions and their impact on sustainability during the design, or re-design. For example, \citet{Fatima2024-sus-cs} perform a sustainability assessment of architecture design decisions. In their dataset, they discuss an example of a \textit{scalability} feature of the cloud application. This feature \texttt{represents} the design decision \texttt{pertaining to} the design concern \textit{availability in peak use time}. The design decision options are \textit{`auto-scaling vs. manual scaling'}. The QAs \textit{`resource utilization, 
availability, and scalability'} \texttt{characterize} the chosen design decision. 


The feature itself is \texttt{realised by} one or more concrete architectural elements.

To provide a glimpse of the potential of the planned integration of software architecture descriptions into the SAF Toolkit, in~\Cref{fig:42010model} we show an example specific for the  domain of service-oriented systems that we use in an educational context\footnote{One of the author is teaching a Master-level course on Service Oriented Design. There, software engineering competencies are taught on sustainability-focused projects~\cite{Condori_Fernandez2018}.} (see dashed classes in~\Cref{fig:42010model}). As such, a software service corresponds to a specific service-oriented system and its components.
If explicitly modeled, such association would allow us to locate architecture elements in the software architecture and create the necessary mapping between features, decisions, and concerns. Only by introducing this mapping can we accurately evaluate the impact  on sustainability of both the taken decisions and the specific architectural elements.

\paragraph{Reflections}
Our goal is to \textbf{\textit{establish traceability}} between architecture elements and their sustainability impacts. We need to understand which elements are responsible for the measured impact and to what extent. Such knowledge will provide us with information on where in the architecture we need to improve. Future direction is needed to frame this traceability as, for example, a reusable catalogue of architecture viewpoints (\ie "conventions for creating, interpreting, presenting and analysing a view"~\cite{ISO_42010_2022}) or model kinds (\ie "conventions for model-based view components"~\cite{ISO_42010_2022}). Another possibility would be to build on a definition of sustainability as a QA and, as such, to frame related concerns as architecture perspectives (\ie "a collection of activities, tactics, and guidelines used to ensure that a system exhibits a particular set of related quality properties that require consideration across a number of the system's architectural views"~\cite{RozanskiWoods_SoftwareSystems_2005}).

It is worth noting that the newly introduced \textbf{\textit{concept of aspects}} in accordance with the Software Architecture ISO/IEC/IEEE 42010:2022 Standard~\cite{ISO_42010_2022} can include concerns or non-functional properties (cf. \Cref{fig:KPImodel}). 
In other words, aspects are the "\textit{visible} evidence (in architecture description models and views) for how concerns are addressed"~\cite{rich}.
However, in the current version of the standard, aspects are only linked to properties of interest but are not explicitly linked to features resulting in a missing relationship to the architecture element. Therefore, the notion of aspects cannot yet be linked to the SAF Toolkit. We identify two challenges: 
(i) how can aspects be linked to features? 
(ii) how can we effectively create the necessary traceability as mentioned above?
Tackling both challenges will help us to successfully provide architects with the necessary feedback about which of the taken decision(s) have an impact on which architecture element(s) and therefore on sustainability.

In the context of the KPI model, \textbf{\textit{actions are triggered}} if a KPI does not reach its target. These actions are derived from the metrics that quantify the KPI. These metrics can be used to identify the QR they represent and the associated design decisions. Hence, we can say that actions can drive the choice of making a certain design decision if sustainability is not achieved over time. However, as mentioned above, an underlying assumption is that one can trace the measured impact to architecture elements.

\subsubsection{Integration with the Green Lab} \label{sec:greenlab}

The Green Lab is the experimental platform developed at Vrije Universiteit Amsterdam for conducting empirical studies on the quality of software, with a special emphasis on its energy efficiency. 
At the time of writing, the Green Lab includes: (i) a cluster of 7 servers with different technical specifications for executing experiments targeting Cloud software, datacenters, and applications executing compute-intensive workloads, such as simulation, machine learning, and scientific computing, (ii) about 20 smartphones and 3 tablets for executing experiments targeting mobile applications, (iii) 32 Raspberry Pi and 8 Arduino Nano boards to test embedded and IoT software, and (iv) other equipment including GPUs, educational robots, VR headsets, and wearables.  
All devices in the Green Lab are equipped with state-of-the-art power monitors/profilers, both hardware and software (\eg Monsoon power monitor, Watts Up Pro Meter, Intel RAPL, INA219).

The Green Lab also provides a set of open-source tools for orchestrating experiments on the energy efficiency of software. Among them, Experiment Runner~\cite{S2_Group_Experiment_Runner} provides a Python library to collect low-level measures from the power monitor/profiler, transform them into higher-level metrics (\eg energy), and persist them according to an open file format (\eg JSON), depending on the needs of the experiment. 

We are planning to integrate (a custom version of) Experiment Runner in the SAF Toolkit. Specifically, by referring to \Cref{fig:KPImodel}, Experiment Runner will be in charge of (i) empirically collecting concrete measures from a running system/prototype (\eg its average power consumption, CPU usage, exchanged HTTP messages) and (ii) making them available within the KPI model in the form of instances of the \texttt{Measure} concept. Collected measures will then be used to enrich the DM related to the running system. 

\paragraph{Reflections}
We expect that software architects using the DM enriched with the collected measures will have a better view of how the system is faring with respect to its KPIs, and thus the reasoning process of the architects will be more precise and grounded in the concrete characteristics of the system at hand. Specifically, the integration of Experiment Runner in the SAF Toolkit can consist in the inclusion of a new step in a continuous integration and continuous deployment (CI/CD) pipeline dedicated to collecting sustainability-related metrics about the system; the newly-added step allows developers to automatically trigger the execution of Experiment Runner for launching a small-scale experiment targeting the system deployed in a testing/pre-production environment. The sustainability-related metrics collected during the experiment are then parsed, aggregated, and suitably mapped into an architectural model of the system, related concepts in the DM, and the impacted KPIs. The main stakeholders benefiting from this enhanced CI/CD pipeline are software architects, as they will be able to continuously have an up-to-date and evidence-based overview of how the various components of the system fare with respect to its overall KPIs and the performed design decisions. We are planning a campaign of user studies involving students and practitioners to either substantiate or disprove the just-mentioned expectation. 


\subsubsection{Integration with Monitoring Dashboard} \label{sec:dashboard}

In this paper, we primarily referred to the SAF Toolkit as a set of tools for supporting the \textit{design} of software-intensive systems with sustainability as a first-class concern.  
However, the interest of software architects and decision makers is not confined to the construction and evolution of a software-intensive system, but rather it is critical to efficiently and reliably \textit{operate} it throughout its complete life time~\cite{fitzgerald2014continuous,van2009continuous}. 
This reflection is not new per se, but, with the advent of the microservices architectural style~\cite{di2019architecting,waseem2021design} and the DevOps~\cite{devops} paradigm, continuously operating a software-intensive system, promptly detecting relevant incidents (\eg faults, performance issues), and diagnosing their potential impact \textit{at runtime} has reached an unprecedented level of importance. 

We believe that sustainability decisions are no exception and that the SAF Toolkit will be a valuable tool for continuously evaluating sustainability decisions (and their direct/indirect impact) at runtime. 
Central to this new line of research is the integration of the SAF Toolkit with a monitoring dashboard. 
In this context, monitoring consists of the collection of runtime metrics (\eg energy consumption, memory/CPU usage, number of DB queries, average service latency) and their analysis to keep the overall quality of the system at acceptable levels~\cite{ebert2016devops}. 
As of today, tens of monitoring tools exist for monitoring systems in the context of DevOps~\cite{JSS_2024}, such as Prometheus\footnote{\url{https://prometheus.io}} or NetData\footnote{\url{https://www.netdata.cloud}}. 
Monitoring platforms, \eg Grafana\footnote{\url{https://grafana.com}} provide dashboarding functionalities for querying the collected measures and graphically visualizing them via interactive tables, histograms, time series graphs, heatmaps, \etc  
Technically, the integration of the SAF Toolkit with a monitoring dashboard will consist of (i) a dedicated connector between the monitoring tool collecting measures at runtime and the dashboard,  (ii) a widget for representing SAF instruments (\eg DMs) in the dashboard, and (iii) a set of mechanisms for mapping the tool-specific collected measures to elements of the KPI model, and in turn to the visualized SAF instrument. 

\paragraph{Reflections}
Using the SAF Toolkit in combination with a monitoring dashboard will be one of the first steps toward the integration of the architecture decision-making process and DevOps. As such, it will be valuable to expand the SAF Toolkit also with functionalities for helping architects promptly understanding the current situation of the system, identifying architecturally-relevant issues (\eg the presence of performance bottlenecks or energy hotspots), and taking timely decisions about how to communicate and solve them. In this context, the usage of large language models tuned for suggesting architectural solutions/tactics in the context of the SAF Toolkit is a promising research direction that we are planning to pursue in future work.

\section{The SAF Toolkit in Action} \label{sec:SAFinAction}

\begin{figure}[H]
    \centering
    \includegraphics[width=0.9\linewidth]{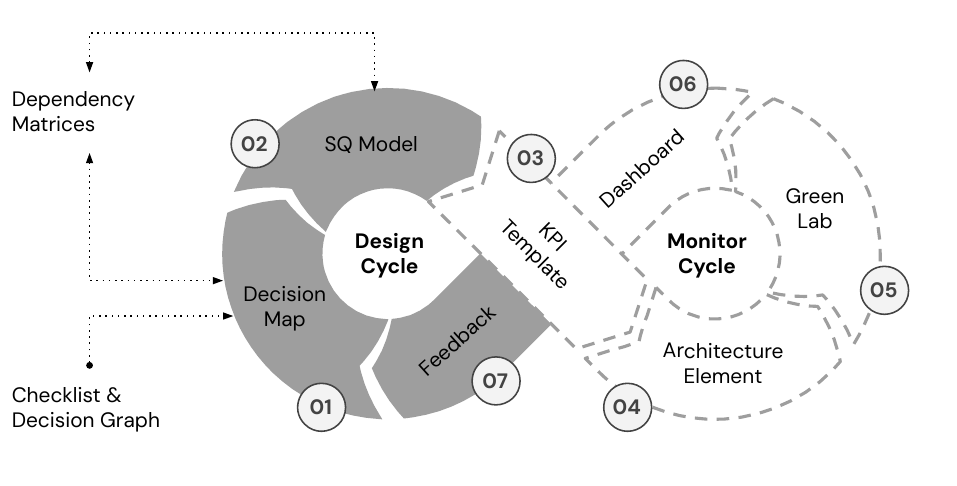}
    \caption{\textbf{SAF Toolkit in Action divided in two vision cycles.} The gray phases in the \textit{Design Cycle} are available in the current version; the dashed phases in the \textit{Monitor Cycle} are part of our future vision and have yet to be integrated.}
    \label{fig:SAFinAction}
\end{figure}

In this section we outline how we envision the complete SAF Toolkit workflow for its intended stakeholders. Architects and decision makers can use this knowledge to understand how the toolkit could be integrated into their current work process; educators can use it as a teaching guide on what steps are necessary to address sustainability holistically in the software engineering process. As shown in \Cref{fig:SAFinAction}, we distinguish between two vision cycles.

The \textbf{Design Cycle} focuses on the incremental design of the DM (01), the SQ model (02) and the use of the KPI template (03).  In this iterative process, the Checklist and Decision Graph (see \Cref{sec:guidingInstrument}) are used as instruments to guide the creation of a comprehensive DM.  
The Dependency Matrices (see \Cref{sec:DMatrix}) support the decision making process by revealing the relationships between the identified QAs in the DM and SQ Model. The matrices can also be revised and extended during the design process to suit the needs of the context and the experience gained. 
We expect the Design Cycle to be completed by formulating KPIs accordingly. The cycle will be re-initiated with evidence-based feedback (07) from the Monitor Cycle.

The \textbf{Monitor Cycle} is concerned with activities that are applied and executed at the system's runtime. In order to facilitate the feedback from the implemented system back to the Design Cycle, it is necessary to establish a mapping of Architecture Elements (04), integrate the Green Lab (05), and employ certain Monitoring Dashboards (06). 
By mapping architecture-relevant features, decisions, and concerns to concrete architecture elements in the software architecture as described in \Cref{sec:SAF+SA}, we can identify which metrics should be measured where.
This understanding is essential for the Green Lab to collect concrete measures from an implemented architecture. Monitoring Dashboards would gather the measured evidence and feed them back to the Design Cycle.

\section{Discussion and Conclusion}\label{sec:discussion}

In this paper, we present the SAF Toolkit, a set of instruments aimed to support architects and design decision makers in modeling sustainability as a software quality property.
In the experience gained over time from the series of case studies, we identified two main types of stakeholders as target users of the toolkit. The first are \textit{practicing architects and decision markers}: in the many cases we could develop with them, we observed that architects consistently found the toolkit useful, both for novices that are new to sustainability in software architecting, and for experienced architects that are already knowledgeable of the relevance of sustainability-quality concerns in their projects but do not know how to extend their practices to apply it. The second type of target stakeholders are \textit{instructors} that want to teach their students about sustainability in software engineering/architecture: we have years of experience in successfully doing that in master level courses, experience that we have translated in the supplementary material~\cite{saf_toolkit_online}, like small video tutorials (work in progress) and the checklist. Of course, practitioners willing to train themselves would find this beneficial, too.

When we described the current release of the SAF Toolkit (in~\Cref{sec:SAFnow}) and the extensions we planned (in \Cref{sec:SAFfuture}), we provided some reflections on the lessons learned and points for improvement. 
After over a decade of research in collaboration with industry and continuous improvements, the two instruments of DM and SQ Model in the current toolkit release are quite mature and consolidated into a powerful yet simple support for practitioners that may be both experienced and novices. Yet, we are missing the two extensions that introduce instruments at design time (defining sound KPIs and linking sustainability elements with architecture elements other than features), and the two extensions that link the toolkit to a measurement environment (like our Green Lab) and a monitoring and reflection environment (like the integration with data collection and dashboarding).

To provide traceability between architecture descriptions (\eg architecture elements in existing views) and the instruments of the SAF Toolkit (\eg quality concerns in a DM and KPIs in an SQ Model), we need further collaboration with industry: most modern architectures will already be dealing with quality concerns; so it is fair to assume that in those cases (i) the SQ Model is extracted from existing architecture descriptions, and (ii) such descriptions comply with established organisational practices. To ease the adoption of the toolkit for adding sustainability to the architecture practice of an organization, we believe that the elements of the SAF Toolkit Overview (see Figure~\ref{fig:structuralModel}) could be used as a pattern that is repeated for each major sustainability concern.

Finally, we also care to go back to the definition of sustainability we have provided in~\Cref{sec:background}, and our emphasis on the indivisible dimensions of \textit{sustainability-quality} and of \textit{sustainability-time}. While time is represented in the DM by means of the mapping on impact levels (immediate, enabling and systemic), such representation is static as a DM is meant to visualize a model in a single point in time. Differently, we can truly assess if a software architecture addresses SQ concerns or is helping achieve a targeted sustainability goal only if we observe (\ie measure and visualize) the trend distributed over a time period. Partially, the integration with the dashboard has the potential to represent measures (hence, effects) over time. This, however, needs major research investments and is a main topic of our future research.

\section*{Acknowledgements}

We would like to thank Rich Hilliard for his thoughtful feedback on an earlier version of this paper. And of course the many practitioners that donated their time, know-how, and experience; and the many students who enthused us for their education: we hope that experimenting with our SAF Toolkit was as valuable for them as it has been for us. 

This work received partial funding from the project SustainableCloud (OCENW.M20.243) of the research programme Open Competition which is (partly) financed by the Dutch Research Council (NWO).

\backmatter

\bibliography{00_MAIN}


\begin{thebibliography}{71}
\ifx \bisbn   \undefined \def \bisbn  #1{ISBN #1}\fi
\ifx \binits  \undefined \def \binits#1{#1}\fi
\ifx \bauthor  \undefined \def \bauthor#1{#1}\fi
\ifx \batitle  \undefined \def \batitle#1{#1}\fi
\ifx \bjtitle  \undefined \def \bjtitle#1{#1}\fi
\ifx \bvolume  \undefined \def \bvolume#1{\textbf{#1}}\fi
\ifx \byear  \undefined \def \byear#1{#1}\fi
\ifx \bissue  \undefined \def \bissue#1{#1}\fi
\ifx \bfpage  \undefined \def \bfpage#1{#1}\fi
\ifx \blpage  \undefined \def \blpage #1{#1}\fi
\ifx \burl  \undefined \def \burl#1{\textsf{#1}}\fi
\ifx \doiurl  \undefined \def \doiurl#1{\url{https://doi.org/#1}}\fi
\ifx \betal  \undefined \def \betal{\textit{et al.}}\fi
\ifx \binstitute  \undefined \def \binstitute#1{#1}\fi
\ifx \binstitutionaled  \undefined \def \binstitutionaled#1{#1}\fi
\ifx \bctitle  \undefined \def \bctitle#1{#1}\fi
\ifx \beditor  \undefined \def \beditor#1{#1}\fi
\ifx \bpublisher  \undefined \def \bpublisher#1{#1}\fi
\ifx \bbtitle  \undefined \def \bbtitle#1{#1}\fi
\ifx \bedition  \undefined \def \bedition#1{#1}\fi
\ifx \bseriesno  \undefined \def \bseriesno#1{#1}\fi
\ifx \blocation  \undefined \def \blocation#1{#1}\fi
\ifx \bsertitle  \undefined \def \bsertitle#1{#1}\fi
\ifx \bsnm \undefined \def \bsnm#1{#1}\fi
\ifx \bsuffix \undefined \def \bsuffix#1{#1}\fi
\ifx \bparticle \undefined \def \bparticle#1{#1}\fi
\ifx \barticle \undefined \def \barticle#1{#1}\fi
\bibcommenthead
\ifx \bconfdate \undefined \def \bconfdate #1{#1}\fi
\ifx \botherref \undefined \def \botherref #1{#1}\fi
\ifx \url \undefined \def \url#1{\textsf{#1}}\fi
\ifx \bchapter \undefined \def \bchapter#1{#1}\fi
\ifx \bbook \undefined \def \bbook#1{#1}\fi
\ifx \bcomment \undefined \def \bcomment#1{#1}\fi
\ifx \oauthor \undefined \def \oauthor#1{#1}\fi
\ifx \citeauthoryear \undefined \def \citeauthoryear#1{#1}\fi
\ifx \endbibitem  \undefined \def \endbibitem {}\fi
\ifx \bconflocation  \undefined \def \bconflocation#1{#1}\fi
\ifx \arxivurl  \undefined \def \arxivurl#1{\textsf{#1}}\fi
\csname PreBibitemsHook\endcsname

\bibitem[\protect\citeauthoryear{{European Commission} and {Directorate-General Energy}}{2023}]{EED2023}
\begin{bbook}
\bauthor{\bsnm{{European Commission}}},
\bauthor{\bsnm{{Directorate-General Energy}}}:
\bbtitle{{Reporting Requirements on the Energy Performance and Sustainability of Data Centres for the Energy Efficiency Directive. Task A Report, Options for a Reporting Scheme for Data Centres}}.
\bpublisher{Publications Office of the European Union},
\blocation{online}
(\byear{2023}).
\burl{https://data.europa.eu/doi/10.2833/304891}
\end{bbook}
\endbibitem

\bibitem[\protect\citeauthoryear{{Techieus}}{2023}]{cloudsoftware2023a}
\begin{botherref}
\oauthor{\bsnm{{Techieus}}}:
{Cloud Computing Adoption: Statistics and Insights Across Sectors}.
\url{https://techieus.com/cloud-computing-adoption-statistics-and-insights-across-sectors}.
Accessed: 18-10-2024
(2023)
\end{botherref}
\endbibitem

\bibitem[\protect\citeauthoryear{Eurostat}{}]{cloudsoftware2023b}
\begin{botherref}
\oauthor{\bsnm{Eurostat}}:
{Cloud computing: statistics on the use by enterprises}.
\url{https://ec.europa.eu/eurostat/statistics-explained/index.php?title=Cloud_computing_-_statistics_on_the_use_by_enterprises}.
Accessed: 01-05-2024
\end{botherref}
\endbibitem

\bibitem[\protect\citeauthoryear{Moghaddam et~al.}{2015}]{Moghaddam2015}
\begin{barticle}
\bauthor{\bsnm{Moghaddam}, \binits{F.A.}},
\bauthor{\bsnm{Lago}, \binits{P.}},
\bauthor{\bsnm{Grosso}, \binits{P.}}:
\batitle{{Energy-Efficient Networking Solutions in Cloud-Based Environments: A Systematic Literature Review}}.
\bjtitle{ACM Comput. Surv.}
\bvolume{47}(\bissue{4}),
\bfpage{1}--\blpage{32}
(\byear{2015})
\end{barticle}
\endbibitem

\bibitem[\protect\citeauthoryear{Gozalo-Brizuela and Garrido-Merchan}{2023}]{generative_ai}
\begin{botherref}
\oauthor{\bsnm{Gozalo-Brizuela}, \binits{R.}},
\oauthor{\bsnm{Garrido-Merchan}, \binits{E.C.}}:
{ChatGPT} is not all you need. A State of the Art Review of large Generative {AI} models
(2023).
\url{https://arxiv.org/abs/2301.04655}
\end{botherref}
\endbibitem

\bibitem[\protect\citeauthoryear{Verdecchia et~al.}{2023}]{verdecchia2023systematic}
\begin{barticle}
\bauthor{\bsnm{Verdecchia}, \binits{R.}},
\bauthor{\bsnm{Sallou}, \binits{J.}},
\bauthor{\bsnm{Cruz}, \binits{L.}}:
\batitle{{A systematic review of Green {AI}}}.
\bjtitle{Wiley interdisciplinary reviews. Data mining and knowledge discovery}
\bvolume{13}(\bissue{4}),
\bfpage{1507}
(\byear{2023})
\end{barticle}
\endbibitem

\bibitem[\protect\citeauthoryear{Lago and {et al.}}{}]{saf_toolkit_online}
\begin{botherref}
\oauthor{\bsnm{Lago}, \binits{P.}}, et al.:
The Sustainability Assessment Framework (SAF) Toolkit.
\url{https://github.com/S2-group/SAF-Toolkit}.
Online Repository.
\end{botherref}
\endbibitem

\bibitem[\protect\citeauthoryear{{Condori-Fernandez} et~al.}{2020}]{Condori-FernandezEtAl_ActionResearch_2020}
\begin{barticle}
\bauthor{\bsnm{{Condori-Fernandez}}, \binits{N.}},
\bauthor{\bsnm{Lago}, \binits{P.}},
\bauthor{\bsnm{Luaces}, \binits{M.R.}},
\bauthor{\bsnm{Places}, \binits{{\'A}.S.}}:
\batitle{An {{Action Research}} for {{Improving}} the {{Sustainability Assessment Framework Instruments}}}.
\bjtitle{Sustainability}
\bvolume{12}(\bissue{4}),
\bfpage{1682}
(\byear{2020})
\end{barticle}
\endbibitem

\bibitem[\protect\citeauthoryear{Calero et~al.}{2020}]{CaleroEtAl_5WsGreen_2020}
\begin{barticle}
\bauthor{\bsnm{Calero}, \binits{C.}},
\bauthor{\bsnm{Mancebo}, \binits{J.}},
\bauthor{\bsnm{Garcia}, \binits{F.}},
\bauthor{\bsnm{Moraga}, \binits{M.A.}},
\bauthor{\bsnm{Berna}, \binits{J.A.G.}},
\bauthor{\bsnm{{Fernandez-Aleman}}, \binits{J.L.}},
\bauthor{\bsnm{Toval}, \binits{A.}}:
\batitle{{{5Ws}} of green and sustainable software}.
\bjtitle{Tsinghua Science and Technology}
\bvolume{25}(\bissue{3}),
\bfpage{401}--\blpage{414}
(\byear{2020})
\end{barticle}
\endbibitem

\bibitem[\protect\citeauthoryear{Peters et~al.}{2024}]{PetersEtAl_SustainabilityComputing_2024}
\begin{botherref}
\oauthor{\bsnm{Peters}, \binits{A.-K.}},
\oauthor{\bsnm{Capilla}, \binits{R.}},
\oauthor{\bsnm{Coroam{\u a}}, \binits{V.C.}},
\oauthor{\bsnm{Heldal}, \binits{R.}},
\oauthor{\bsnm{Lago}, \binits{P.}},
\oauthor{\bsnm{Leifler}, \binits{O.}},
\oauthor{\bsnm{Moreira}, \binits{A.}},
\oauthor{\bsnm{Fernandes}, \binits{J.P.}},
\oauthor{\bsnm{Penzenstadler}, \binits{B.}},
\oauthor{\bsnm{Porras}, \binits{J.}},
\oauthor{\bsnm{Venters}, \binits{C.C.}}:
Sustainability in {{Computing Education}}: {{A Systematic Literature Review}}.
ACM Transactions on Computing Education,
3639060
(2024)
\end{botherref}
\endbibitem

\bibitem[\protect\citeauthoryear{Calero and Piattini}{2017}]{CaleroPiattini_PuzzlingOut_2017}
\begin{barticle}
\bauthor{\bsnm{Calero}, \binits{C.}},
\bauthor{\bsnm{Piattini}, \binits{M.}}:
\batitle{Puzzling out {{Software Sustainability}}}.
\bjtitle{Sustainable Computing: Informatics and Systems}
\bvolume{16},
\bfpage{117}--\blpage{124}
(\byear{2017})
\end{barticle}
\endbibitem

\bibitem[\protect\citeauthoryear{Penzenstadler}{2013}]{Penzenstadler2013-3criteria}
\begin{bchapter}
\bauthor{\bsnm{Penzenstadler}, \binits{B.}}:
\bctitle{{Towards a definition of sustainabilityinandforsoftware engineering}}.
In: \bbtitle{{Proceedings of the 28th Annual ACM Symposium on Applied Computing(SAC)}}.
\bpublisher{ACM Press},
\blocation{New York, NY, USA}
(\byear{2013})
\end{bchapter}
\endbibitem

\bibitem[\protect\citeauthoryear{Wolfram et~al.}{2017}]{Wolfram2017-def}
\begin{bchapter}
\bauthor{\bsnm{Wolfram}, \binits{N.}},
\bauthor{\bsnm{Lago}, \binits{P.}},
\bauthor{\bsnm{Osborne}, \binits{F.}}:
\bctitle{{Sustainability in software engineering}}.
In: \bbtitle{{Sustainable Internet and ICT for Sustainability (SustainIT)}},
pp. \bfpage{1}--\blpage{7}.
\bpublisher{{IEEE}},
\blocation{Funchal, Portugal}
(\byear{2017}).
\burl{http://dx.doi.org/10.23919/SustainIT.2017.8379798}
\end{bchapter}
\endbibitem

\bibitem[\protect\citeauthoryear{Venters et~al.}{2023}]{VentersEtAl_SustainableSoftware_2023}
\begin{botherref}
\oauthor{\bsnm{Venters}, \binits{C.C.}},
\oauthor{\bsnm{Capilla}, \binits{R.}},
\oauthor{\bsnm{Nakagawa}, \binits{E.Y.}},
\oauthor{\bsnm{Betz}, \binits{S.}},
\oauthor{\bsnm{Penzenstadler}, \binits{B.}},
\oauthor{\bsnm{Crick}, \binits{T.}},
\oauthor{\bsnm{Brooks}, \binits{I.}}:
Sustainable software engineering: {{Reflections}} on advances in research and practice.
Information and Software Technology,
107316
(2023)
\end{botherref}
\endbibitem

\bibitem[\protect\citeauthoryear{Lago et~al.}{2015}]{LagoEtAl_FramingSustainability_2015}
\begin{barticle}
\bauthor{\bsnm{Lago}, \binits{P.}},
\bauthor{\bsnm{Ko{\c c}ak}, \binits{S.A.}},
\bauthor{\bsnm{Crnkovic}, \binits{I.}},
\bauthor{\bsnm{Penzenstadler}, \binits{B.}}:
\batitle{Framing sustainability as a property of software quality}.
\bjtitle{Communications of the ACM}
\bvolume{58}(\bissue{10}),
\bfpage{70}--\blpage{78}
(\byear{2015})
\end{barticle}
\endbibitem

\bibitem[\protect\citeauthoryear{Lago}{2023}]{Lago_ConnectedWorld_2023}
\begin{bchapter}
\bauthor{\bsnm{Lago}, \binits{P.}}:
\bctitle{{The Digital Society Is Already Here -- Pity It Is 'Unsustainable'}}.
In: \beditor{\bsnm{Vermeulen}, \binits{I.}} (ed.)
\bbtitle{{Connected World - Insights from 100 Academics on How to Build Better Connections}},
pp. \bfpage{55}--\blpage{58}.
\bpublisher{VU University Press},
\blocation{Amsterdam}
(\byear{2023}).
\burl{https://vuuniversitypress.com/product/connected-world}
\end{bchapter}
\endbibitem

\bibitem[\protect\citeauthoryear{Becker et~al.}{2015}]{BeckerEtAl_SustainabilityDesign_2015}
\begin{bchapter}
\bauthor{\bsnm{Becker}, \binits{C.}},
\bauthor{\bsnm{Chitchyan}, \binits{R.}},
\bauthor{\bsnm{Duboc}, \binits{L.}},
\bauthor{\bsnm{Easterbrook}, \binits{S.}},
\bauthor{\bsnm{Penzenstadler}, \binits{B.}},
\bauthor{\bsnm{Seyff}, \binits{N.}},
\bauthor{\bsnm{Venters}, \binits{C.C.}}:
\bctitle{Sustainability {{Design}} and {{Software}}: {{The Karlskrona Manifesto}}}.
In: \bbtitle{2015 {{IEEE}}/{{ACM}} 37th {{IEEE International Conference}} on {{Software Engineering}}},
pp. \bfpage{467}--\blpage{476}.
\bpublisher{{IEEE}},
\blocation{{Florence, Italy}}
(\byear{2015})
\end{bchapter}
\endbibitem

\bibitem[\protect\citeauthoryear{Venters et~al.}{2014}]{VentersEtAl_BlindMen_2014}
\begin{botherref}
\oauthor{\bsnm{Venters}, \binits{C.C.}},
\oauthor{\bsnm{Lau}, \binits{L.}},
\oauthor{\bsnm{Griffiths}, \binits{M.K.}},
\oauthor{\bsnm{Holmes}, \binits{V.}},
\oauthor{\bsnm{Ward}, \binits{R.R.}},
\oauthor{\bsnm{Jay}, \binits{C.}},
\oauthor{\bsnm{Dibsdale}, \binits{C.E.}},
\oauthor{\bsnm{Xu}, \binits{J.}}:
The {{Blind Men}} and the {{Elephant}}: {{Towards}} an {{Empirical Evaluation Framework}} for {{Software Sustainability}}.
Journal of Open Research Software
\textbf{2}(1)
(2014)
\end{botherref}
\endbibitem

\bibitem[\protect\citeauthoryear{Calero et~al.}{2021}]{CaleroEtAl_IntroductionSoftware_2021}
\begin{bchapter}
\bauthor{\bsnm{Calero}, \binits{C.}},
\bauthor{\bsnm{Moraga}, \binits{M.{\'A}.}},
\bauthor{\bsnm{Piattini}, \binits{M.}}:
\bctitle{Introduction to {{Software Sustainability}}}.
In: \beditor{\bsnm{Calero}, \binits{C.}},
\beditor{\bsnm{Moraga}, \binits{M.{\'A}.}},
\beditor{\bsnm{Piattini}, \binits{M.}} (eds.)
\bbtitle{Software {{Sustainability}}},
pp. \bfpage{1}--\blpage{15}.
\bpublisher{{Springer International Publishing}},
\blocation{{Cham}}
(\byear{2021})
\end{bchapter}
\endbibitem

\bibitem[\protect\citeauthoryear{Andrikopoulos et~al.}{2022}]{AndrikopoulosEtAl_SustainabilitySoftware_2022}
\begin{bchapter}
\bauthor{\bsnm{Andrikopoulos}, \binits{V.}},
\bauthor{\bsnm{Boza}, \binits{R.-D.}},
\bauthor{\bsnm{Perales}, \binits{C.}},
\bauthor{\bsnm{Lago}, \binits{P.}}:
\bctitle{Sustainability in {{Software Architecture}}: {{A Systematic Mapping Study}}}.
In: \bbtitle{2022 48th {{Euromicro Conference}} on {{Software Engineering}} and {{Advanced Applications}} ({{SEAA}})},
pp. \bfpage{426}--\blpage{433}.
\bpublisher{{IEEE}},
\blocation{{Gran Canaria, Spain}}
(\byear{2022})
\end{bchapter}
\endbibitem

\bibitem[\protect\citeauthoryear{Heldal et~al.}{2024}]{Heldal2024-industry}
\begin{barticle}
\bauthor{\bsnm{Heldal}, \binits{R.}},
\bauthor{\bsnm{Nguyen}, \binits{N.-T.}},
\bauthor{\bsnm{Moreira}, \binits{A.}},
\bauthor{\bsnm{Lago}, \binits{P.}},
\bauthor{\bsnm{Duboc}, \binits{L.}},
\bauthor{\bsnm{Betz}, \binits{S.}},
\bauthor{\bsnm{Coroam{\u a}}, \binits{V.C.}},
\bauthor{\bsnm{Penzenstadler}, \binits{B.}},
\bauthor{\bsnm{Porras}, \binits{J.}},
\bauthor{\bsnm{Capilla}, \binits{R.}},
\bauthor{\bsnm{Brooks}, \binits{I.}},
\bauthor{\bsnm{Oyedeji}, \binits{S.}},
\bauthor{\bsnm{Venters}, \binits{C.C.}}:
\batitle{{Sustainability competencies and skills in software engineering: An industry perspective}}.
\bjtitle{The Journal of systems and software}
\bvolume{211},
\bfpage{111978}
(\byear{2024})
\end{barticle}
\endbibitem

\bibitem[\protect\citeauthoryear{Nandonde}{2019}]{Nandonde_PESTLEAnalysis_2019}
\begin{barticle}
\bauthor{\bsnm{Nandonde}, \binits{F.A.}}:
\batitle{A {{PESTLE}} analysis of international retailing in the {{East African Community}}}.
\bjtitle{Global Business and Organizational Excellence}
\bvolume{38}(\bissue{4}),
\bfpage{54}--\blpage{61}
(\byear{2019})
\end{barticle}
\endbibitem

\bibitem[\protect\citeauthoryear{{European Banking Authority}}{2022}]{EuropeanBankingAuthority_ReportIncorporating_2022}
\begin{botherref}
\oauthor{\bsnm{{European Banking Authority}}}:
Report on incorporating {{ESG}} risks in the supervision of investment firms.
Technical Report EBA/REP/2022/26
(2022)
\end{botherref}
\endbibitem

\bibitem[\protect\citeauthoryear{Duboc et~al.}{2020}]{DubocEtAl_RequirementsEngineering_2020}
\begin{barticle}
\bauthor{\bsnm{Duboc}, \binits{L.}},
\bauthor{\bsnm{Penzenstadler}, \binits{B.}},
\bauthor{\bsnm{Porras}, \binits{J.}},
\bauthor{\bsnm{Akinli~Kocak}, \binits{S.}},
\bauthor{\bsnm{Betz}, \binits{S.}},
\bauthor{\bsnm{Chitchyan}, \binits{R.}},
\bauthor{\bsnm{Leifler}, \binits{O.}},
\bauthor{\bsnm{Seyff}, \binits{N.}},
\bauthor{\bsnm{Venters}, \binits{C.C.}}:
\batitle{Requirements engineering for sustainability: An awareness framework for designing software systems for a better tomorrow}.
\bjtitle{Requirements Engineering}
\bvolume{25}(\bissue{4}),
\bfpage{469}--\blpage{492}
(\byear{2020})
\end{barticle}
\endbibitem

\bibitem[\protect\citeauthoryear{Hilty and Aebischer}{2015}]{Hilty_2015}
\begin{bchapter}
\bauthor{\bsnm{Hilty}, \binits{L.M.}},
\bauthor{\bsnm{Aebischer}, \binits{B.}}:
\bctitle{{ICT} for sustainability: An emerging research field}.
In: \bbtitle{Advances in Intelligent Systems and Computing}.
\bsertitle{Advances in intelligent systems and computing},
pp. \bfpage{3}--\blpage{36}.
\bpublisher{Springer},
\blocation{Cham}
(\byear{2015})
\end{bchapter}
\endbibitem

\bibitem[\protect\citeauthoryear{Parmenter}{2015}]{Parmenter_2015}
\begin{bbook}
\bauthor{\bsnm{Parmenter}, \binits{D.}}:
\bbtitle{Key Performance Indicators: Developing, Implementing, and Using Winning KPIs}.
\bpublisher{John Wiley \& Sons},
\blocation{UK}
(\byear{2015})
\end{bbook}
\endbibitem

\bibitem[\protect\citeauthoryear{Fatima et~al.}{2024}]{Fatima_2023}
\begin{botherref}
\oauthor{\bsnm{Fatima}, \binits{I.}},
\oauthor{\bsnm{Funke}, \binits{M.}},
\oauthor{\bsnm{Lago}, \binits{P.}}:
Providing guidance to software practitioners: A framework for creating {KPIs}.
IEEE Software,
1--9
(2024)
\doiurl{10.1109/MS.2024.3456446} .
In Press.
\end{botherref}
\endbibitem

\bibitem[\protect\citeauthoryear{Staron et~al.}{2016}]{Staron_2016}
\begin{bchapter}
\bauthor{\bsnm{Staron}, \binits{M.}},
\bauthor{\bsnm{Meding}, \binits{W.}},
\bauthor{\bsnm{Niesel}, \binits{K.}},
\bauthor{\bsnm{Abran}, \binits{A.}}:
\bctitle{A key performance indicator quality model and its industrial evaluation}.
In: \bbtitle{2016 Joint Conference of the International Workshop on Software Measurement and the International Conference on Software Process and Product Measurement (IWSM-MENSURA)},
pp. \bfpage{170}--\blpage{179}
(\byear{2016}).
\doiurl{10.1109/IWSM-Mensura.2016.033}
\end{bchapter}
\endbibitem

\bibitem[\protect\citeauthoryear{Staron et~al.}{2012}]{Staron_2012}
\begin{bchapter}
\bauthor{\bsnm{Staron}, \binits{M.}},
\bauthor{\bsnm{Meding}, \binits{W.}},
\bauthor{\bsnm{Palm}, \binits{K.}}:
\bctitle{Release readiness indicator for mature agile and lean software development projects}.
In: \bbtitle{Lecture Notes in Business Information Processing}.
\bsertitle{Lecture notes in business information processing},
pp. \bfpage{93}--\blpage{107}.
\bpublisher{Springer},
\blocation{Berlin, Heidelberg}
(\byear{2012})
\end{bchapter}
\endbibitem

\bibitem[\protect\citeauthoryear{Naumann et~al.}{2011}]{Naumann_2011}
\begin{barticle}
\bauthor{\bsnm{Naumann}, \binits{S.}},
\bauthor{\bsnm{Dick}, \binits{M.}},
\bauthor{\bsnm{Kern}, \binits{E.}},
\bauthor{\bsnm{Johann}, \binits{T.}}:
\batitle{The greensoft model: A reference model for green and sustainable software and its engineering}.
\bjtitle{Sustainable Computing: Informatics and Systems}
\bvolume{1}(\bissue{4}),
\bfpage{294}--\blpage{304}
(\byear{2011})
\doiurl{10.1016/j.suscom.2011.06.004}
\end{barticle}
\endbibitem

\bibitem[\protect\citeauthoryear{Cabot et~al.}{2009}]{Cabot_2009}
\begin{bchapter}
\bauthor{\bsnm{Cabot}, \binits{J.}},
\bauthor{\bsnm{Easterbrook}, \binits{S.}},
\bauthor{\bsnm{Horkoff}, \binits{J.}},
\bauthor{\bsnm{Lessard}, \binits{L.}},
\bauthor{\bsnm{Liaskos}, \binits{S.}},
\bauthor{\bsnm{Mazon}, \binits{J.-N.}}:
\bctitle{Integrating sustainability in decision-making processes: A modelling strategy}.
In: \bbtitle{31st International Conference on Software Engineering - Companion Volume},
pp. \bfpage{207}--\blpage{210}
(\byear{2009}).
\doiurl{10.1109/ICSE-COMPANION.2009.5070983}
\end{bchapter}
\endbibitem

\bibitem[\protect\citeauthoryear{Penzenstadler and Femmer}{2013}]{Penzenstadler_2013}
\begin{bchapter}
\bauthor{\bsnm{Penzenstadler}, \binits{B.}},
\bauthor{\bsnm{Femmer}, \binits{H.}}:
\bctitle{A generic model for sustainability with process- and product-specific instances}.
In: \bbtitle{Proceedings of the Workshop on Green In/by Software Engineering}.
\bsertitle{GIBSE},
pp. \bfpage{3}--\blpage{8}.
\bpublisher{ACM},
\blocation{NY, USA}
(\byear{2013}).
\doiurl{10.1145/2451605.2451609}
\end{bchapter}
\endbibitem

\bibitem[\protect\citeauthoryear{Saputri and Lee}{2016}]{Saputri_2016}
\begin{bchapter}
\bauthor{\bsnm{Saputri}, \binits{T.R.D.}},
\bauthor{\bsnm{Lee}, \binits{S.-W.}}:
\bctitle{Incorporating sustainability design in requirements engineering process: A preliminary study}.
In: \beditor{\bsnm{Lee}, \binits{S.-W.}},
\beditor{\bsnm{Nakatani}, \binits{T.}} (eds.)
\bbtitle{Requirements Engineering Toward Sustainable World},
pp. \bfpage{53}--\blpage{67}.
\bpublisher{Springer},
\blocation{Singapore}
(\byear{2016})
\end{bchapter}
\endbibitem

\bibitem[\protect\citeauthoryear{Duboc et~al.}{2020}]{Duboc_2020}
\begin{barticle}
\bauthor{\bsnm{Duboc}, \binits{L.}},
\bauthor{\bsnm{Penzenstadler}, \binits{B.}},
\bauthor{\bsnm{Porras}, \binits{J.}},
\bauthor{\bsnm{Akinli~Kocak}, \binits{S.}},
\bauthor{\bsnm{Betz}, \binits{S.}},
\bauthor{\bsnm{Chitchyan}, \binits{R.}},
\bauthor{\bsnm{Leifler}, \binits{O.}},
\bauthor{\bsnm{Seyff}, \binits{N.}},
\bauthor{\bsnm{Venters}, \binits{C.C.}}:
\batitle{{Requirements engineering for sustainability: an awareness framework for designing software systems for a better tomorrow}}.
\bjtitle{Requirements Engineering}
\bvolume{25}(\bissue{4}),
\bfpage{469}--\blpage{492}
(\byear{2020})
\doiurl{10.1007/s00766-020-00336-y}
\end{barticle}
\endbibitem

\bibitem[\protect\citeauthoryear{Penzenstadler et~al.}{2023}]{SusAF_Vision_2023}
\begin{bchapter}
\bauthor{\bsnm{Penzenstadler}, \binits{B.}},
\bauthor{\bsnm{Seyff}, \binits{N.}},
\bauthor{\bsnm{Betz}, \binits{S.}},
\bauthor{\bsnm{Duboc}, \binits{L.}},
\bauthor{\bsnm{Porras}, \binits{J.}},
\bauthor{\bsnm{Chitchyan}, \binits{R.}},
\bauthor{\bsnm{Brooks}, \binits{I.}},
\bauthor{\bsnm{Oyedeji}, \binits{S.}},
\bauthor{\bsnm{Villela}, \binits{K.B.}},
\bauthor{\bsnm{Venters}, \binits{C.C.}}:
\bctitle{{Vision Paper: The Sustainability Awareness Framework (SusAF) as a De-Facto Standard?}}
In: \bbtitle{First International Workshop on Requirements Engineering Frameworks ({REFrame}) at {REFSQ}},
vol. \bseriesno{3378},
p. \bfpage{4}
(\byear{2023}).
\bcomment{CEUR Workshop Proceedings}
\end{bchapter}
\endbibitem

\bibitem[\protect\citeauthoryear{}{}]{GAISSA}
\begin{botherref}
Towards green AI‐based software systems: an architecture‐centric approach. GAISSA.
\url{https://gaissa.upc.edu/en}.
[Accessed 18-10-2024]
\end{botherref}
\endbibitem

\bibitem[\protect\citeauthoryear{Guldner et~al.}{2024}]{GuldnerEtAl_DevelopmentEvaluation_2024a}
\begin{barticle}
\bauthor{\bsnm{Guldner}, \binits{A.}},
\bauthor{\bsnm{Bender}, \binits{R.}},
\bauthor{\bsnm{Calero}, \binits{C.}},
\bauthor{\bsnm{Fernando}, \binits{G.S.}},
\bauthor{\bsnm{Funke}, \binits{M.}},
\bauthor{\bsnm{Gr{\"o}ger}, \binits{J.}},
\bauthor{\bsnm{Hilty}, \binits{L.M.}},
\bauthor{\bsnm{H{\"o}rnschemeyer}, \binits{J.}},
\bauthor{\bsnm{Hoffmann}, \binits{G.-D.}},
\bauthor{\bsnm{Junger}, \binits{D.}},
\bauthor{\bsnm{Kennes}, \binits{T.}},
\bauthor{\bsnm{Kreten}, \binits{S.}},
\bauthor{\bsnm{Lago}, \binits{P.}},
\bauthor{\bsnm{Mai}, \binits{F.}},
\bauthor{\bsnm{Malavolta}, \binits{I.}},
\bauthor{\bsnm{Murach}, \binits{J.}},
\bauthor{\bsnm{Oberg{\"o}ker}, \binits{K.}},
\bauthor{\bsnm{Schmidt}, \binits{B.}},
\bauthor{\bsnm{Tarara}, \binits{A.}},
\bauthor{\bsnm{{De Veaugh-Geiss}}, \binits{J.P.}},
\bauthor{\bsnm{Weber}, \binits{S.}},
\bauthor{\bsnm{Westing}, \binits{M.}},
\bauthor{\bsnm{Wohlgemuth}, \binits{V.}},
\bauthor{\bsnm{Naumann}, \binits{S.}}:
\batitle{Development and evaluation of a reference measurement model for assessing the resource and energy efficiency of software products and components---{{Green Software Measurement Model}} ({{GSMM}})}.
\bjtitle{Future Generation Computer Systems}
\bvolume{155},
\bfpage{402}--\blpage{418}
(\byear{2024})
\end{barticle}
\endbibitem

\bibitem[\protect\citeauthoryear{Fatima and Lago}{2024}]{Fatima2024-sus-cs}
\begin{bchapter}
\bauthor{\bsnm{Fatima}, \binits{I.}},
\bauthor{\bsnm{Lago}, \binits{P.}}:
\bctitle{{Software Architecture Assessment for Sustainability: A Case Study}}.
In: \beditor{\bsnm{{Galster, Matthias and Scandurra, Patrizia and Mikkonen, Tommi and Oliveira Antonino, Pablo and Nakagawa, Elisa Yumi and Navarro, Elena}}} (ed.)
\bbtitle{{European Conference on Software Architecture}}.
\bsertitle{Lecture Notes in Computer Science},
pp. \bfpage{233}--\blpage{249}.
\bpublisher{Springer},
\blocation{Switzerland}
(\byear{2024})
\end{bchapter}
\endbibitem

\bibitem[\protect\citeauthoryear{{Object Manamgement Group}}{2017}]{OMG-UML251}
\begin{botherref}
\oauthor{\bsnm{{Object Manamgement Group}}}:
{OMG unified modeling language (UML) version 2.5.1}
(2017).
\url{https://www.omg.org/spec/UML}
\end{botherref}
\endbibitem

\bibitem[\protect\citeauthoryear{de~Boer and van Vliet}{2009}]{boer_vanvliet2009}
\begin{barticle}
\bauthor{\bsnm{Boer}, \binits{R.C.}},
\bauthor{\bsnm{Vliet}, \binits{H.}}:
\batitle{On the similarity between requirements and architecture}.
\bjtitle{Journal of Systems and Software}
\bvolume{82}(\bissue{3}),
\bfpage{544}--\blpage{550}
(\byear{2009})
\doiurl{10.1016/j.jss.2008.11.185}
\end{barticle}
\endbibitem

\bibitem[\protect\citeauthoryear{{International Organization for Standardization [ISO]}}{2011}]{ISO_25010_2011}
\begin{botherref}
\oauthor{\bsnm{{International Organization for Standardization [ISO]}}}:
Systems and software engineering - {{Systems}} and software {{Quality Requirements}} and {{Evaluation}} ({{SQuaRE}}) - {{System}} and software quality models.
Technical Report ISO/IEC 25010:2011
(2011)
\end{botherref}
\endbibitem

\bibitem[\protect\citeauthoryear{{Condori-Fernandez} and Lago}{2018}]{Condori-FernandezLago_CharacterizingContribution_2018}
\begin{barticle}
\bauthor{\bsnm{{Condori-Fernandez}}, \binits{N.}},
\bauthor{\bsnm{Lago}, \binits{P.}}:
\batitle{Characterizing the contribution of quality requirements to software sustainability}.
\bjtitle{Journal of Systems and Software}
\bvolume{137},
\bfpage{289}--\blpage{305}
(\byear{2018})
\end{barticle}
\endbibitem

\bibitem[\protect\citeauthoryear{Betz et~al.}{2015}]{BetzEtAl_SustainabilityDebt_2015}
\begin{bchapter}
\bauthor{\bsnm{Betz}, \binits{S.}},
\bauthor{\bsnm{Becker}, \binits{C.}},
\bauthor{\bsnm{Chitchyan}, \binits{R.}},
\bauthor{\bsnm{Duboc}, \binits{L.}},
\bauthor{\bsnm{Easterbrook}, \binits{S.M.}},
\bauthor{\bsnm{Penzenstadler}, \binits{B.}},
\bauthor{\bsnm{Seyff}, \binits{N.}},
\bauthor{\bsnm{Venters}, \binits{C.C.}}:
\bctitle{Sustainability {{Debt}}: {{A Metaphor}} to {{Support Sustainability Design Decisions}}}.
In: \bbtitle{Fourth {{International Workshop}} on {{Requirements Engineering}} for {{Sustainable Systems}} ({{RE4SuSy}})},
\bconflocation{{Ottawa, Canada}}
(\byear{2015})
\end{bchapter}
\endbibitem

\bibitem[\protect\citeauthoryear{}{2022}]{ISO_42010_2022}
\begin{botherref}
{ISO/IEC/IEEE 42010, Software, Systems and Enterprise --- Architecture Description}.
(2022)
\end{botherref}
\endbibitem

\bibitem[\protect\citeauthoryear{Salama et~al.}{2021}]{Salama2021}
\begin{barticle}
\bauthor{\bsnm{Salama}, \binits{M.}},
\bauthor{\bsnm{Bahsoon}, \binits{R.}},
\bauthor{\bsnm{Lago}, \binits{P.}}:
\batitle{{Stability in software engineering: Survey of the state-of-the-art and research directions}}.
\bjtitle{{IEEE} Transactions on Software Engineering}
\bvolume{47}(\bissue{7}),
\bfpage{1468}--\blpage{1510}
(\byear{2021})
\end{barticle}
\endbibitem

\bibitem[\protect\citeauthoryear{Funke et~al.}{2023}]{FunkeEtAl_VariabilityFeatures_2023}
\begin{bchapter}
\bauthor{\bsnm{Funke}, \binits{M.}},
\bauthor{\bsnm{Lago}, \binits{P.}},
\bauthor{\bsnm{Verdecchia}, \binits{R.}}:
\bctitle{Variability {{Features}}: {{Extending Sustainability Decision Maps}} via an {{Industrial Case Study}}}.
In: \bbtitle{2023 {{IEEE}} 20th {{International Conference}} on {{Software Architecture Companion}} ({{ICSA-C}})},
pp. \bfpage{1}--\blpage{7}.
\bpublisher{{IEEE}},
\blocation{{L'Aquila, Italy}}
(\byear{2023})
\end{bchapter}
\endbibitem

\bibitem[\protect\citeauthoryear{Wohlin and Rainer}{2022}]{Wohlin2022-CS-checklist}
\begin{barticle}
\bauthor{\bsnm{Wohlin}, \binits{C.}},
\bauthor{\bsnm{Rainer}, \binits{A.}}:
\batitle{{Is it a case study?—A critical analysis and guidance}}.
\bjtitle{The Journal of systems and software}
\bvolume{192}(\bissue{111395}),
\bfpage{111395}
(\byear{2022})
\doiurl{10.1016/j.jss.2022.111395}
\end{barticle}
\endbibitem

\bibitem[\protect\citeauthoryear{Espana and Lago}{2016}]{Sergio_Espana2016}
\begin{botherref}
\oauthor{\bsnm{Espana}, \binits{S.}},
\oauthor{\bsnm{Lago}, \binits{P.}}:
{Software Sustainability Assessment (SoSA) exercise report}.
Technical report,
Vrije Universiteit Amsterdam
(May 2016).
\url{https://goo.gl/d9FYi9}
\end{botherref}
\endbibitem

\bibitem[\protect\citeauthoryear{Verdecchia et~al.}{2017}]{Verdecchia2017}
\begin{bchapter}
\bauthor{\bsnm{Verdecchia}, \binits{R.}},
\bauthor{\bsnm{Procaccianti}, \binits{G.}},
\bauthor{\bsnm{Malavolta}, \binits{I.}},
\bauthor{\bsnm{Lago}, \binits{P.}},
\bauthor{\bsnm{Koedijk}, \binits{J.}}:
\bctitle{{Estimating Energy Impact of Software Releases and Deployment Strategies: The KPMG Case Study}}.
In: \beditor{\bsnm{Nurcan}, \binits{S.}},
\beditor{\bsnm{Soffer}, \binits{P.}},
\beditor{\bsnm{Bajec}, \binits{M.}},
\beditor{\bsnm{Eder}, \binits{J.}} (eds.)
\bbtitle{International Symposium on Empirical Software Engineering and Measurement (ESEM)},
pp. \bfpage{257}--\blpage{266}.
\bpublisher{Springer},
\blocation{Ljubljana, Slovenia}
(\byear{2017}).
\doiurl{10.1109/ESEM.2017.39}
\end{bchapter}
\endbibitem

\bibitem[\protect\citeauthoryear{Niggebrugge et~al.}{2018}]{Niggebrugge2018}
\begin{botherref}
\oauthor{\bsnm{Niggebrugge}, \binits{T.}},
\oauthor{\bsnm{Vos}, \binits{S.}},
\oauthor{\bsnm{Lago}, \binits{P.}}:
{The Sustainability of Mobility as a Service Solutions Evaluated through the Software Sustainability Assessment Method}.
Technical report,
Vrije Universiteit Amsterdam
(January 2018)
\end{botherref}
\endbibitem

\bibitem[\protect\citeauthoryear{Lago}{2019}]{Lago2019}
\begin{bchapter}
\bauthor{\bsnm{Lago}, \binits{P.}}:
\bctitle{{Architecture Design Decision Maps for Software Sustainability}}.
In: \bbtitle{41st International Conference on Software Engineering: Software Engineering in Society (ICSE-SEIS)}.
\bsertitle{ICSE},
pp. \bfpage{61}--\blpage{64}.
\bpublisher{IEEE/ACM},
\blocation{Montr{\'e}al, Canada}
(\byear{2019}).
\burl{http://dx.doi.org/10.1109/ICSE-SEIS.2019.00015}
\end{bchapter}
\endbibitem

\bibitem[\protect\citeauthoryear{Vos et~al.}{2020a}]{Vos2020-tr}
\begin{botherref}
\oauthor{\bsnm{Vos}, \binits{S.}},
\oauthor{\bsnm{Schaefers}, \binits{H.}},
\oauthor{\bsnm{Bon}, \binits{A.}},
\oauthor{\bsnm{Lago}, \binits{P.}}:
{Sustainability and Ethics by Design: On the development of digital platforms in low-resource environments}.
Technical report,
Vrije Universiteit Amsterdam
(December 2020).
\url{https://research.vu.nl/files/137357217/ICT4FoodSec_1_.pdf}
\end{botherref}
\endbibitem

\bibitem[\protect\citeauthoryear{Vos et~al.}{2020b}]{Vos2020-paper}
\begin{bchapter}
\bauthor{\bsnm{Vos}, \binits{S.}},
\bauthor{\bsnm{Schaefers}, \binits{H.}},
\bauthor{\bsnm{Lago}, \binits{P.}},
\bauthor{\bsnm{Bon}, \binits{A.}}:
\bctitle{{Towards Sustainability and Equality in Digital Development}}.
In: \beditor{\bsnm{{Bandi, R.K., C R, R., Klein, S., Madon, S., Monteiro, E.}}} (ed.)
\bbtitle{The Future of Digital Work: The Challenge of Inequality}.
\bsertitle{IFIP AICT},
vol. \bseriesno{601}.
\bpublisher{Springer},
\blocation{Cham}
(\byear{2020}).
\burl{http://link.springer.com.vu-nl.idm.oclc.org/978-3-030-64697-4}
\end{bchapter}
\endbibitem

\bibitem[\protect\citeauthoryear{Bischoff et~al.}{2022}]{Bischoff2021}
\begin{bchapter}
\bauthor{\bsnm{Bischoff}, \binits{Y.}},
\bauthor{\bsnm{Wiel}, \binits{R.}},
\bauthor{\bsnm{Hooff}, \binits{B.}},
\bauthor{\bsnm{Lago}, \binits{P.}}:
\bctitle{A taxonomy about information systems complexity and sustainability}.
In: \beditor{\bsnm{Wohlgemuth}, \binits{V.}},
\beditor{\bsnm{Naumann}, \binits{S.}},
\beditor{\bsnm{Behrens}, \binits{G.}},
\beditor{\bsnm{Arndt}, \binits{H.-K.}} (eds.)
\bbtitle{Advances and New Trends in Environmental Informatics},
pp. \bfpage{17}--\blpage{33}.
\bpublisher{Springer},
\blocation{Cham}
(\byear{2022})
\end{bchapter}
\endbibitem

\bibitem[\protect\citeauthoryear{Lago et~al.}{2021}]{Lago2021-winterschool}
\begin{bchapter}
\bauthor{\bsnm{Lago}, \binits{P.}},
\bauthor{\bsnm{Verdecchia}, \binits{R.}},
\bauthor{\bsnm{Condori-Fernandez}, \binits{N.}},
\bauthor{\bsnm{Rahmadian}, \binits{E.}},
\bauthor{\bsnm{Sturm}, \binits{J.}},
\bauthor{\bsnm{Nijnanten}, \binits{T.}},
\bauthor{\bsnm{Bosma}, \binits{R.}},
\bauthor{\bsnm{Debuysscher}, \binits{C.}},
\bauthor{\bsnm{Ricardo}, \binits{P.}}:
\bctitle{{Designing for Sustainability: Lessons Learned from Four Industrial Projects}}.
In: \bbtitle{{Advances and New Trends in Environmental Informatics}},
pp. \bfpage{3}--\blpage{18}.
\bpublisher{Springer},
\blocation{Cham}
(\byear{2021}).
\doiurl{10.1007/978-3-030-61969-5_1}
\end{bchapter}
\endbibitem

\bibitem[\protect\citeauthoryear{Condori-Fernandez et~al.}{2024}]{CondoriFernandez2024-interdepend}
\begin{botherref}
\oauthor{\bsnm{Condori-Fernandez}, \binits{N.}},
\oauthor{\bsnm{Lago}, \binits{P.}},
\oauthor{\bsnm{Catala}, \binits{A.}},
\oauthor{\bsnm{Luaces}, \binits{M.R.}}:
{Defining Interdimensional Dependencies of the Sustainability-Quality Model}.
Technical report,
Vrije Universiteit Amsterdam
(March 2024)
\end{botherref}
\endbibitem

\bibitem[\protect\citeauthoryear{Quispe-Cruz and Condori-Fernandez}{2022}]{CIBSE22}
\begin{bchapter}
\bauthor{\bsnm{Quispe-Cruz}, \binits{M.}},
\bauthor{\bsnm{Condori-Fernandez}, \binits{N.}}:
\bctitle{Extending the sustainability-quality model for supporting the design of persuasive software systems}.
In: \bbtitle{Anais do XXV Congresso Ibero-Americano em Engenharia de Software},
pp. \bfpage{158}--\blpage{172}.
\bpublisher{SBC},
\blocation{Porto Alegre, RS, Brasil}
(\byear{2022}).
\doiurl{10.5753/cibse.2022.20970}
\end{bchapter}
\endbibitem

\bibitem[\protect\citeauthoryear{Checkland}{1999}]{checkland1999systems}
\begin{botherref}
\oauthor{\bsnm{Checkland}, \binits{P.}}:
Systems thinking. rethinking management information systems.
Rethinking: Management information systems: An interdisciplinary perspective,
44--56
(1999)
\end{botherref}
\endbibitem

\bibitem[\protect\citeauthoryear{Deckers}{2023}]{deckers}
\begin{botherref}
\oauthor{\bsnm{Deckers}, \binits{R.}}
Personal Communication.
(2023)
\end{botherref}
\endbibitem

\bibitem[\protect\citeauthoryear{Apel and K{\"a}stner}{2009}]{Apel2009}
\begin{barticle}
\bauthor{\bsnm{Apel}, \binits{S.}},
\bauthor{\bsnm{K{\"a}stner}, \binits{C.}}:
\batitle{{An Overview of Feature-Oriented Software Development}}.
\bjtitle{Journal of Object Technology}
\bvolume{8}(\bissue{5}),
\bfpage{49}--\blpage{84}
(\byear{2009})
\end{barticle}
\endbibitem

\bibitem[\protect\citeauthoryear{Condori~Fernandez and Lago}{2018}]{Condori_Fernandez2018}
\begin{bchapter}
\bauthor{\bsnm{Condori~Fernandez}, \binits{N.}},
\bauthor{\bsnm{Lago}, \binits{P.}}:
\bctitle{{The Influence of Green Strategies Design onto Quality Requirements Prioritization}}.
In: \bbtitle{{Requirements Engineering: Foundation for Software Quality}},
pp. \bfpage{189}--\blpage{205}.
\bpublisher{Springer},
\blocation{Cham}
(\byear{2018}).
\burl{http://dx.doi.org/10.1007/978-3-319-77243-1_12}
\end{bchapter}
\endbibitem

\bibitem[\protect\citeauthoryear{Rozanski and Woods}{2005}]{RozanskiWoods_SoftwareSystems_2005}
\begin{bbook}
\bauthor{\bsnm{Rozanski}, \binits{N.}},
\bauthor{\bsnm{Woods}, \binits{E.}}:
\bbtitle{Software Systems Architecture: Working with Stakeholders Using Viewpoints and Perspectives}.
\bpublisher{Addison-Wesley},
\blocation{Upper Saddle River, NJ}
(\byear{2005})
\end{bbook}
\endbibitem

\bibitem[\protect\citeauthoryear{Hilliard}{2024}]{rich}
\begin{botherref}
\oauthor{\bsnm{Hilliard}, \binits{R.}}
Personal Communication.
(2024)
\end{botherref}
\endbibitem

\bibitem[\protect\citeauthoryear{{Software and Sustanability (S2) Group}}{}]{S2_Group_Experiment_Runner}
\begin{botherref}
\oauthor{\bsnm{{Software and Sustanability (S2) Group}}}:
{Experiment Runner}.
\url{https://github.com/s2-group/experiment-runner/}
\end{botherref}
\endbibitem

\bibitem[\protect\citeauthoryear{Fitzgerald and Stol}{2014}]{fitzgerald2014continuous}
\begin{bchapter}
\bauthor{\bsnm{Fitzgerald}, \binits{B.}},
\bauthor{\bsnm{Stol}, \binits{K.-J.}}:
\bctitle{Continuous software engineering and beyond: trends and challenges}.
In: \bbtitle{Proceedings of the 1st International Workshop on Rapid Continuous Software Engineering},
pp. \bfpage{1}--\blpage{9}
(\byear{2014})
\end{bchapter}
\endbibitem

\bibitem[\protect\citeauthoryear{van Hoorn et~al.}{2009}]{van2009continuous}
\begin{botherref}
\oauthor{\bsnm{Hoorn}, \binits{A.}},
\oauthor{\bsnm{Hasselbring}, \binits{W.}},
\oauthor{\bsnm{Waller}, \binits{J.}},
\oauthor{\bsnm{Ehlers}, \binits{J.}},
\oauthor{\bsnm{Frey}, \binits{S.}},
\oauthor{\bsnm{Kieselhorst}, \binits{D.}}:
Continuous monitoring of software services: Design and application of the kieker framework
(2009)
\end{botherref}
\endbibitem

\bibitem[\protect\citeauthoryear{Di~Francesco et~al.}{2019}]{di2019architecting}
\begin{barticle}
\bauthor{\bsnm{Di~Francesco}, \binits{P.}},
\bauthor{\bsnm{Lago}, \binits{P.}},
\bauthor{\bsnm{Malavolta}, \binits{I.}}:
\batitle{Architecting with microservices: A systematic mapping study}.
\bjtitle{Journal of Systems and Software}
\bvolume{150},
\bfpage{77}--\blpage{97}
(\byear{2019})
\end{barticle}
\endbibitem

\bibitem[\protect\citeauthoryear{Waseem et~al.}{2021}]{waseem2021design}
\begin{barticle}
\bauthor{\bsnm{Waseem}, \binits{M.}},
\bauthor{\bsnm{Liang}, \binits{P.}},
\bauthor{\bsnm{Shahin}, \binits{M.}},
\bauthor{\bsnm{Di~Salle}, \binits{A.}},
\bauthor{\bsnm{M{\'a}rquez}, \binits{G.}}:
\batitle{Design, monitoring, and testing of microservices systems: The practitioners’ perspective}.
\bjtitle{Journal of Systems and Software}
\bvolume{182},
\bfpage{111061}
(\byear{2021})
\end{barticle}
\endbibitem

\bibitem[\protect\citeauthoryear{Jabbari et~al.}{2016}]{devops}
\begin{bchapter}
\bauthor{\bsnm{Jabbari}, \binits{R.}},
\bauthor{\bsnm{Ali}, \binits{N.}},
\bauthor{\bsnm{Petersen}, \binits{K.}},
\bauthor{\bsnm{Tanveer}, \binits{B.}}:
\bctitle{What is {DevOps}? a systematic mapping study on definitions and practices}.
In: \bbtitle{Proceedings of the Scientific Workshop Proceedings of XP2016}.
\bsertitle{XP '16 Workshops}.
\bpublisher{Association for Computing Machinery},
\blocation{New York, NY, USA}
(\byear{2016}).
\doiurl{10.1145/2962695.2962707}
\end{bchapter}
\endbibitem

\bibitem[\protect\citeauthoryear{Ebert et~al.}{2016}]{ebert2016devops}
\begin{barticle}
\bauthor{\bsnm{Ebert}, \binits{C.}},
\bauthor{\bsnm{Gallardo}, \binits{G.}},
\bauthor{\bsnm{Hernantes}, \binits{J.}},
\bauthor{\bsnm{Serrano}, \binits{N.}}:
\batitle{{DevOps}}.
\bjtitle{IEEE Software}
\bvolume{33}(\bissue{3}),
\bfpage{94}--\blpage{100}
(\byear{2016})
\doiurl{10.1109/MS.2016.68}
\end{barticle}
\endbibitem

\bibitem[\protect\citeauthoryear{Giamattei et~al.}{2024}]{JSS_2024}
\begin{barticle}
\bauthor{\bsnm{Giamattei}, \binits{L.}},
\bauthor{\bsnm{Guerriero}, \binits{A.}},
\bauthor{\bsnm{Pietrantuono}, \binits{R.}},
\bauthor{\bsnm{Russo}, \binits{S.}},
\bauthor{\bsnm{Malavolta}, \binits{I.}},
\bauthor{\bsnm{Islam}, \binits{T.}},
\bauthor{\bsnm{Dinga}, \binits{M.}},
\bauthor{\bsnm{Koziolek}, \binits{A.}},
\bauthor{\bsnm{Singh}, \binits{S.}},
\bauthor{\bsnm{Armbruster}, \binits{M.}},
\bauthor{\bsnm{{Gutierrez-Martinez}}, \binits{J.}},
\bauthor{\bsnm{{Caro-Alvaro}}, \binits{S.}},
\bauthor{\bsnm{Rodriguez}, \binits{D.}},
\bauthor{\bsnm{Weber}, \binits{S.}},
\bauthor{\bsnm{Henss}, \binits{J.}},
\bauthor{\bsnm{{Fernandez Vogelin}}, \binits{E.}},
\bauthor{\bsnm{{Simon Panojo}}, \binits{F.}}:
\batitle{Monitoring tools for devops and microservices: A systematic grey literature review}.
\bjtitle{Journal of Systems and Software}
\bvolume{208},
\bfpage{111906}
(\byear{2024})
\end{barticle}
\endbibitem

\end{thebibliography}

\newpage
\textbf{Patricia Lago} {\,}is a professor of software engineering in the Software and Sustainability research group and the Director of the Digital Sustainability Center at Vrije Universiteit Amsterdam, The Netherlands. Her research focuses primarily on software architecture design and decision making, software quality assessment, and software sustainability. She is the recipient of an Honorary Doctorate at NTNU, Norway, for her contribution to the field of software sustainability, and of the 2023 IEEE CS TCSE New Directions Award. She has a PhD in Control and Computer Engineering from Politecnico di Torino and a Master in Computer Science from the University of Pisa, both in Italy. She is an Associate Editor of IEEE Software, the Elsevier Journal of Systems and Software, and an Advisory Board member of the ACM Transactions on Autonomous and Adaptive Systems. She is a member of ACM, VERSEN, and the Dutch Coalition Sustainable Digitalization (NCDD), and a senior member of IEEE.

\textbf{Nelly Condori-Fernández} {\,}is an assistant professor at the Universidade da Santiago de Compostela (Spain) and a senior researcher of the Centro Singular de Investigación en Tecnoloxías Intelixentes (CiTIUS). 
Her main research focuses on software engineering and human-computer interaction, with a particular interest in leveraging emotion recognition technologies to enhance continuous software improvement. Additionally, her research interests include software sustainability assessment, emphasizing the social and technical aspects of software-intensive systems. She holds a PhD degree in Computer Science from the Universidad Politécnica de Valencia. 

\textbf{Iffat Fatima} {\,}is a PhD candidate with the Software and Sustainability group (S2) at Vrije Universiteit Amsterdam, The Netherlands. She is working on the SustainableCloud project funded by the Dutch Resaerch Council (NWO). Her research interests lie at the intersection of software architecture and sustainability. She holds a master's degree in Software Engineering from the National University of Sciences and Technology in Pakistan.

\textbf{Markus Funke} {\,}is a PhD candidate with the Software and Sustainability group (S2) at Vrije Universiteit Amsterdam, The Netherlands. His research interests include software architecture, architecture knowledge, and software sustainability, with particular interest in the connection to professional practice and the industry. He received his joint Master's degree in Computer Science from the Vrije Universiteit Amsterdam and the University of Amsterdam, The Netherlands. He is a researcher at the VU Digital Sustainability Center (DiSC), the Sustainable IT Lab, and the EU research project µDevOps.

\textbf{Ivano Malavolta} {\,}is an Associate Professor in the Software and Sustainability research group and Director of the Network Institute at the Vrije Universiteit Amsterdam, The Netherlands. His research interests include empirical software engineering, with a special emphasis on software architecture, robotics software, and energy-efficient software. He authored more than 150 scientific articles in peer-reviewed international journals and international conference proceedings. He is program committee member and reviewer of international conferences and journals and Associate Editor of IEEE Software, the International Journal of Robotics Research, and the Frontiers in Robotics and AI journal. He received a PhD in computer science from the University of L’Aquila, Italy. He is a Member of IEEE, ACM, VERSEN, and Amsterdam Data Science.



\end{document}